\documentclass[
twocolumn,
reprint,
superscriptaddress,
nofootinbib,
nobibnotes,
amsmath,
amssymb,
aps,
pra,
longbibliography,
showkeys
]{revtex4-1}
 
\usepackage[latin1]{inputenc}
\usepackage{amsmath}
\usepackage{amsfonts}
\usepackage{hyperref}
\usepackage{amssymb}
\usepackage{graphicx}
\usepackage{hyperref}
\usepackage{colortbl}
\usepackage{subfigure}
\usepackage{amsthm}
\usepackage{thmtools}
\usepackage[normalem]{ulem}
\usepackage{dsfont}
\usepackage{circuitikz}
\usepackage{cases}

\usepackage[T1]{fontenc}
\usepackage{mathptmx}
\usepackage{bbold}
\usepackage{xcolor}

\usepackage{mathtools}

\newcommand{\hyref}[1]{\hyperref[#1]{\ref{#1}}}
\newcommand{\dd}{\mathrm{d}}

\newcommand{\orange}[1]

\renewcommand{\thesection}{\arabic{section}}

\newcommand{\thenewsubsection}{\thesection.\arabic{subsection}}

\makeatletter
\def\@hangfrom@section#1#2#3{\@hangfrom{#1#2}#3}
\def\@hangfroms@section#1#2{#1#2}
\makeatother

\usepackage{titlesec}

\AtBeginDocument{%
    \newwrite\bibnotes
    \def\bibnotesext{Notes.bib}
    \immediate\openout\bibnotes=\jobname\bibnotesext
    \immediate\write\bibnotes{@CONTROL{REVTEX41Control}}
    \immediate\write\bibnotes{@CONTROL{%
    apsrev41Control,author="08",editor="1",pages="1",title="0",year="1"}}
     \if@filesw
     \immediate\write\@auxout{\string\citation{apsrev41Control}}%
    \fi
}%

\titleformat{\section}  
  {\fontsize{12}{14}\bfseries} 
  {\thesection} 
  {1em} 
  {\centering} 
  [] 

\titleformat{\subsection}  
  {\fontsize{10}{12}\bfseries} 
  {\thenewsubsection} 
  {2em} 
  {\centering} 
  [] 
\begin{document}

\title{Advanced iontronic spiking modes with multiscale diffusive dynamics in a fluidic circuit}

\author{T. M. Kamsma}
\thanks{Email: t.m.kamsma@uu.nl; r.vanroij@uu.nl}
\affiliation{Institute for Theoretical Physics, Utrecht University,  Princetonplein 5, 3584 CC Utrecht, The Netherlands}
\affiliation{Mathematical Institute, Utrecht University, Budapestlaan 6, 3584 CD Utrecht, The Netherlands}
\author{E. A. Rossing}
\affiliation{Institute for Theoretical Physics, Utrecht University,  Princetonplein 5, 3584 CC Utrecht, The Netherlands}
\author{C. Spitoni}
\affiliation{Mathematical Institute, Utrecht University, Budapestlaan 6, 3584 CD Utrecht, The Netherlands}
\author{R. van Roij}
\thanks{Email: t.m.kamsma@uu.nl; r.vanroij@uu.nl}
\affiliation{Institute for Theoretical Physics, Utrecht University,  Princetonplein 5, 3584 CC Utrecht, The Netherlands}

\date{\today}
\begin{abstract}
Fluidic iontronics is emerging as a distinctive platform for implementing neuromorphic circuits, characterized by its reliance on the same aqueous medium and ionic signal carriers as the brain. Drawing upon recent theoretical advancements in both iontronic spiking circuits and in dynamic conductance of conical ion channels, which form fluidic memristors, we expand the repertoire of proposed neuronal spiking dynamics in iontronic circuits. Through a modelled circuit containing channels that carry a bipolar surface charge, we extract phasic bursting, mixed-mode spiking, tonic bursting, and threshold variability, all with spike voltages and frequencies within the typical range for mammalian neurons. These features are possible due to the strong dependence of the typical conductance memory retention time on the channel length, enabling timescales varying from individual spikes to bursts of multiple spikes within a single circuit. These advanced forms of neuronal-like spiking support the exploration of aqueous iontronics as an interesting platform for neuromorphic circuits.
\end{abstract}
\keywords{Iontronic spiking, ion channel memristor, microfluidic circuit, tonic bursting, phasic bursting, mixed mode spiking, threshold variability}

\maketitle
\section{Introduction}
In the pursuit of brain-inspired circuits the focus is often on the synaptic properties of neuromorphic devices, where synapses are considered as primary computational units in neuromorphic computing \cite{Schuman2022OpportunitiesApplications}. Consequently, due to their analogous behaviour to synapses, memristors have significantly shaped and driven research in this domain, where the time- and history-dependent conductance of memristors offers a versatile platform for emulating features of synaptic plasticity \cite{Sangwan2020NeuromorphicMaterials,Schuman2017AHardware,Zhu2020ADevices}. However, synapses are not the only components in the brain which can be emulated with memristors. The biological ion channels responsible for generating action potentials also exhibit memristive behavior \cite{Sah2014BrainsMemristors}. This is underscored by the seminal Hodgkin-Huxley (HH) model \cite{Hodgkin1952ANerve}, which mathematically describes the axonal membrane potential by treating the membrane as an equivalent electric circuit in which the ion channels embedded in the axonal membrane are modelled as circuit components. The mathematical models for these ion channels were later recognised as descriptions of memristors \cite{Chua2012Hodgkin-HuxleyMemristors}. Although both synapses and axonal ion channels are neuronal components that can be described and emulated by memristors, they are explicitly distinct biological structures which carry out different tasks. This biological nuance sometimes leads to confusion and inaccurate descriptions of memristive devices in the brain, such as incorrectly associating the HH model with descriptions of synapses \cite{Caravelli2018MemristorsOutsiders}. Nevertheless, the intriguing connection between memristors and the HH model has also sparked considerable interest \cite{Sah2014BrainsMemristors,Chua2013MemristorChaos} and neuronal signalling has inspired various circuits that capture various features of neuronal spiking \cite{Thakur2018Large-ScaleBrain,Yang2020NeuromorphicSystems}. 

Biological neurons feature a wealth of different spiking modes, which can be clearly categorised and used to judge the quality of neuron models \cite{Izhikevich2004WhichNeurons}. Typically the most basic features to consider are \textit{tonic spiking}, a regular train of voltage spikes with constant frequency, and \textit{phasic spiking}, a single isolated voltage spike. In the case of phasic spiking, the neuron model should also obey the all-or-none law \cite{Zhu2020ADevices,Bean2007TheNeurons}, i.e.\ a voltage spike is either fully generated upon a sufficiently strong impulse, or the voltage fails to spike, with no intermediate transition in between. However, many more neuronal firing modes are recognised and this signalling behaviour of neurons has inspired various circuits that can emulate a wide array of different modes of neuronal spiking \cite{Thakur2018Large-ScaleBrain,Yang2020NeuromorphicSystems}. Examples, that will also feature in the present study, include \textit{phasic bursting}, \textit{mixed mode} spiking, \textit{tonic bursting} (otherwise known as \textit{chattering} \cite{Gray1996ChatteringCortex}), and \textit{threshold variability} \cite{Izhikevich2004WhichNeurons}. In phasic bursting, a single burst of several spikes emerges upon applying a sustained stimulus, after which the system again settles to a steady state, despite the constant and sustained current stimulus. Mixed mode spiking consists of an initial burst of spikes upon a sustained stimulus, followed by tonic spiking. In tonic bursting, short periods of spiking, i.e.\ bursts, are interchanged by short periods of no spiking at all. Lastly, threshold variability indicates that the threshold for a neuron to spike can depend on the prior activity of the neuron.
\begin{figure*}[ht]
		\centering
		\includegraphics[width=1\textwidth]{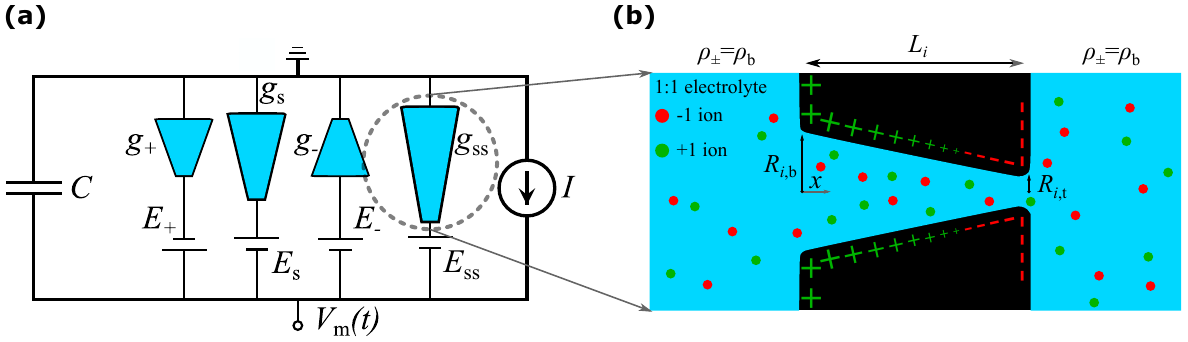}
		\caption{\textbf{(a)} Schematic representation of the proposed fluidic iontronic circuit featuring four channels of three different types. Two short channels of equal length $L_{\pm}=1\text{ }\mu\text{m}$ with fast dynamics on a typical timescale $\tau_{\pm}\approx 0.042$ ms and conductances $g_{\pm}(t)$, a longer channel of length $L_{\mathrm{s}}=15\text{ }\mu\text{m}$ with slower dynamics on a typical timescale $\tau_{\mathrm{s}}\approx 9.4$ ms and conductance $g_{\mathrm{s}}(t)$, and an even longer channel of length $L_{\mathrm{ss}}=90\text{ }\mu\text{m}$ with conductance $g_{\mathrm{ss}}(t)$ and the slowest dynamics over a typical timescale $\tau_{\mathrm{ss}}\approx 338$ ms. These channels are connected in series with batteries with potential $E_{\pm}=\pm114$ mV and $E_{\mathrm{s}}=E_{\mathrm{ss}}=-180$ mV, respectively, and in parallel to a capacitor of capacitance $C=0.05$ pF. A time-dependent stimulus current $I(t)$ can be imposed through the circuit and a potential $V_{\mathrm{m}}(t)$ forms over the circuit that is equivalent to the neuronal membrane potential \cite{Hodgkin1952ANerve}. Schematic adapted from Ref.~\cite{Kamsma2023IontronicMemristors}.
        \textbf{(b)} Schematic of an individual bipolar channel of length $L_i$, with base radius $R_{i,\mathrm{b}}$ and tip radius $R_{i,\mathrm{t}}$, connecting two aqueous 1:1 electrolyte reservoirs of concentration $\rho_{\mathrm{b}}=2$ mM. The wall of all four channels carries an inhomogeneous surface charge that linearly decreases from $0.1\;e\text{nm}^{-2}$ at the base to $-0.05\;e\text{nm}^{-2}$ at the tip.}
		\label{fig:System}
\end{figure*}

The vast majority of neuromorphic devices (both spiking and synaptic) consist (at least partially)  of solid-state components \cite{Thakur2018Large-ScaleBrain,Yang2020NeuromorphicSystems,Sangwan2020NeuromorphicMaterials,Schuman2017AHardware}, which results in fundamental differences with biological neurons. For instance, while solid-state devices typically rely on a single information carrier, such as electrons or holes, driven only by electric forces, neurons employ the transport of various ions and molecules in parallel, while combining electrical and chemical regulation, both for signalling \cite{Micu2017Axo-myelinicSystem} and for synaptic transmission \cite{Pereda2014ElectricalSynapses,Xia2005ThePlasticity,Luscher2012NMDALTP/LTD}. Additionally, the fast dynamics of solid state components can be a disadvantage when temporal inputs are natural or biological signals as the typical timescales of those inputs can be significantly slower than those of solid-state devices, therefore requiring complicated virtual clocks for synchronisation \cite{Covi2021AdaptiveDevices,Chicca2020ASystems}. Recent work tries to address and overcome these limitations through electrochemical coupling of solid-state components to ionic systems, both in the context of synaptic devices \cite{Wang2022DynamicBehaviour,VanDeBurgt2018OrganicComputing} and for spiking circuits \cite{Harikesh2022OrganicSpiking,Harikesh2023IonTunable,Luo2023HighlyDynamics}. However, a newly emerging direction proposes to omit solid-state components altogether, and hence the need for any chemical or ionic coupling, by implementing neuromorphic features in an aqueous electrolyte medium \cite{Noy2023NanofluidicSplash,Noy2023FluidDevices,Robin2021ModelingSlits,Robin2023Long-termChannels,Xiong2023NeuromorphicMemristor,Emmerich2023IonicSwitches,Han2023IontronicDissolution,Xie2022PerspectiveApplication}. These (fluidic) iontronic devices have recently garnered significant interest, offering the promise of multiple information carriers, chemical regulation, and bio-integrability \cite{Han2022Iontronics:Applications}, although sacrificing on the high speeds obtainable by solid state devices. Unlike traditional solid-state neuromorphic circuits, fluidic iontronic circuits leverage the dynamic interplay of ions within an aqueous electrolyte, mirroring the conductive and fluidic characteristics inherent in biological neuronal environments. This departure from solid-state components introduces a novel dimension to neuromorphic computing, offering the potential for closer emulation of the brain's aqueous dynamics \cite{Bocquet2023ConcludingMemories,Xiong2023FluidicDevices}. Recent advances include chemical regulation \cite{Robin2023Long-termChannels,Xiong2023NeuromorphicMemristor} and initial demonstrations of iontronic neuromorphic computing \cite{Kamsma2023Brain-inspiredNanochannels}. However, the development of neuromorphic iontronic devices is still in its infancy, requiring further theoretical explorations and experimental investigations to establish their capabilities in emulating complex neuronal functionalities \cite{Han2022Iontronics:Applications,Xie2022PerspectiveApplication,Noy2023FluidDevices}.

In the recent rise of interest in iontronic neuromorphics, spiking circuits also received some attention in the form of theoretical studies, where HH-inspired iontronic circuits are modelled and shown to exhibit features of neuronal spiking \cite{Robin2021ModelingSlits,Kamsma2023IontronicMemristors}. These proposals feature a circuit composed of an aqueous electrolyte medium, akin to the neuronal medium that the HH model describes, and rely on fluidic iontronic memristors to induce neuronal spiking. Initially, tonic spiking was shown to emerge from a circuit containing angstrom-scale slits \cite{Robin2021ModelingSlits}, shortly after which an alternative iontronic circuit exploiting conical ion channels was proposed that exhibits both the characteristic all-or-none phasic spiking and tonic spiking \cite{Kamsma2023IontronicMemristors}. Thus, the two modes that are typically considered first \cite{Izhikevich2004WhichNeurons,Yang2020NeuromorphicSystems} have been theoretically predicted to also emerge from fluidic iontronic circuits. However, no proposals yet exist to also include other spiking modes.

In this work we expand upon the previously reported features of neuronal spiking in fluidic iontronics \cite{Robin2021ModelingSlits,Kamsma2023IontronicMemristors}. By building upon a previously reported iontronic circuit \cite{Kamsma2023IontronicMemristors} and a physical description of the dynamical conductance of conical channels with a bipolar surface charge \cite{Kamsma2023UnveilingIontronics}, i.e.\ positive at the base and negative at the tip, we can unlock various new forms of spiking dynamics. Due to the strong dependence of the typical conductance memory retention time on the channel length, we can implement timescales varying from individual spikes to bursts of multiple spikes  within a single circuit, thereby enabling new spiking modes. Specifically these spiking modes are the aforementioned phasic bursting, mixed mode spiking, tonic bursting, and threshold variability \cite{Izhikevich2004WhichNeurons}.

\section{Iontronic circuit and bipolar channels}\label{sec:circuit}
Conical fluidic ion channels act as iontronic volatile memristors \cite{Wang2012TransmembraneTransport} and are being investigated as possible candidates for synaptic devices \cite{Ramirez2023SynapticalMemristors} and spiking circuits \cite{Kamsma2023IontronicMemristors,Kamsma2023UnveilingIontronics}. Using theoretical models that quantitatively explain the memristive behaviour of conical channels, it was shown that HH-inspired fluidic circuits containing three conical channels and a capacitor exhibit tonic and phasic spiking \cite{Kamsma2023IontronicMemristors,Kamsma2023UnveilingIontronics}. This modelled circuit was originally composed of conical ion channels with a homogeneous unipolar (UP) surface charge \cite{Kamsma2023IontronicMemristors} and was later modified by replacing the UP channels with conical channels carrying a bipolar (BP) inhomogeneous surface charge \cite{Kamsma2023UnveilingIontronics}, positive at the base and negative at the tip. BP fluidic channels and (Janus) membranes have long drawn great interest as current rectifiers \cite{Daiguji2005NanofluidicTransistor,Vlassiouk2007NanofluidicDiode,Strathmann1997LimitingMembranes,MontesDeOca2022IonicSimulation,Cordoba2023CurrentNanopores,Huang2018BioinspiredBipolar} for applications in e.g.\ sensing \cite{Huang2018BioinspiredBipolar} and osmotic energy conversion \cite{Huang2018BioinspiredBipolar,Yang2018JanusEfficiency,Yan2021PorousFunctionality}. BP channels also show potential for iontronic memristors as a modification from a UP to a BP surface charge led, for an individual conical channel, to a much more pronounced current-voltage hysteresis loop upon applying an AC voltage, i.e.\ a stronger conductance memory effect. 

Here we consider a circuit containing several of these conical BP channels, with different lengths $L_i$. An important feature of these BP channel memristors is that their typical conductance memory timescale is dictated by the channel lengths $L_i$ according to
\begin{align}\label{eq:ts}
    \tau_i=\frac{L_i^2}{12D},
\end{align}
with $D=2\text{ }\mu\text{m}^2\text{ms}^{-1}$ the diffusion coefficient of the ions \cite{Kamsma2023UnveilingIontronics}, which we assume to be identical for all ionic species for convenience. As we will discuss in Sec.~\ref{sec:channelrole}, the combination of channels of various lengths in a single circuit gives rise to dynamics on the timescale of individual spikes and of bursts of spikes.

\begin{figure*}[ht]
		\centering
		\includegraphics[width=1\textwidth]{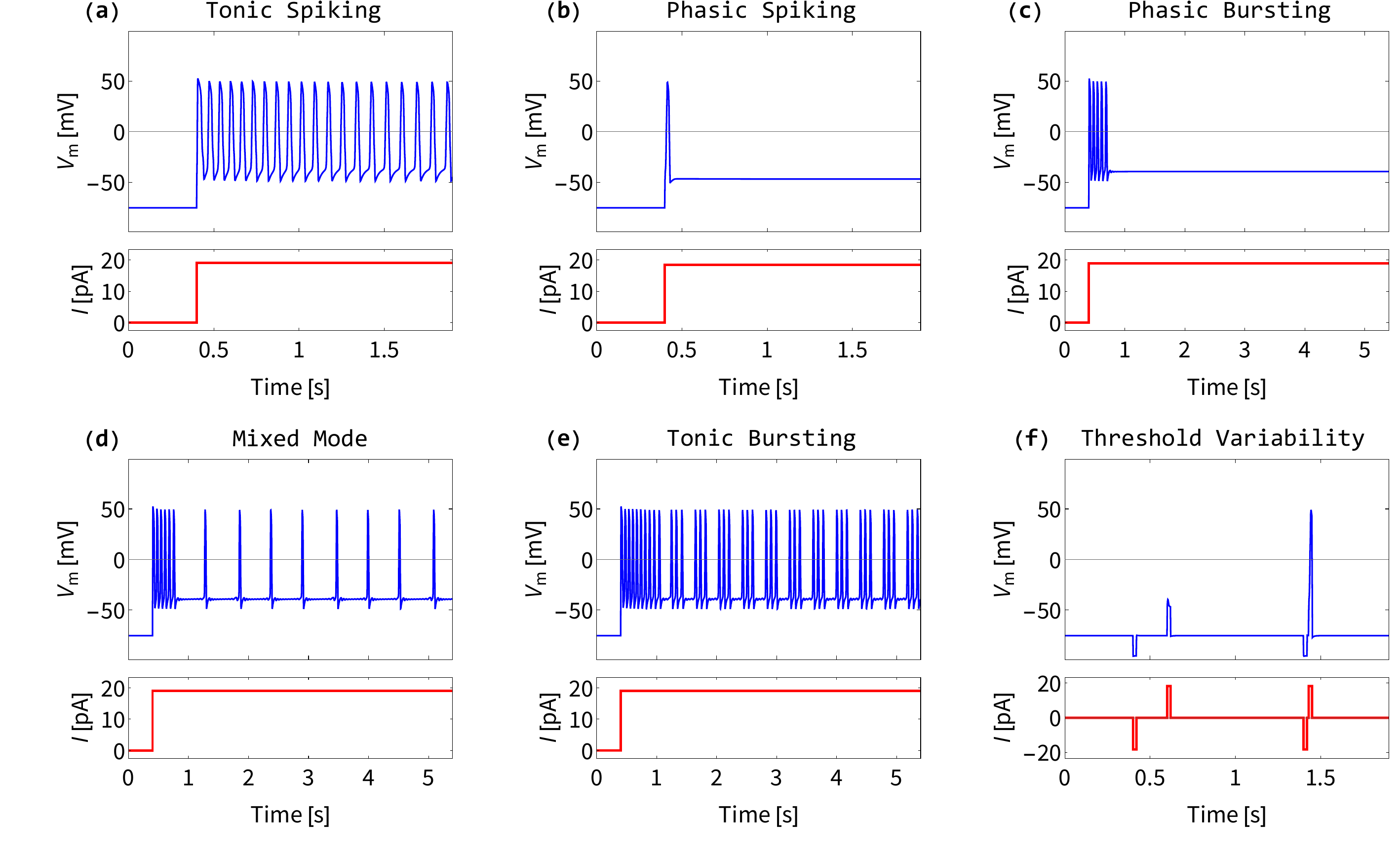}
		\caption{Various modes of voltage spiking (blue curves) extracted by modeling one and the same iontronic circuit driven by different time-dependent currents (red curves), with \textbf{(a)} tonic spiking \cite{Robin2021ModelingSlits,Kamsma2023IontronicMemristors} for $I=19.05$ pA and \textbf{(b)} phasic spiking \cite{Kamsma2023IontronicMemristors} for $I=18.4$ pA reported before in iontronic circuits. The newly introduced modes of iontronic spiking dynamics include \textbf{(c)} phasic bursting for $I=19.01$ pA, i.e.\ a burst of spikes followed by a return to a steady state upon a sustained stimulus, \textbf{(d)} mixed-mode spiking for $I=19.02$ pA, i.e.\ an initial high-frequency burst of spikes followed by a transition into lower frequency tonic spiking, \textbf{(e)} tonic bursting for $I=19.04$ pA,  i.e.\ a short burst of spiking alternating with periods of quiescence, and \textbf{(f)} threshold variability, with variations in the firing threshold influenced by prior activity. The negative and positive current stimuli are of the same magnitude for $I=18.3$ pA but with different time intervals between the negative and the subsequent positive pulse. The firing threshold is temporarily lowered by the negative pulse and therefore the positive pulse only surpasses the (variable) firing threshold when the time between the current pulses is sufficiently short.}
		\label{fig:SpikeGrid}
\end{figure*}

To unlock additional features of neuronal firing, beyond tonic and phasic spiking, we introduce the circuit schematically depicted in Fig.~\ref{fig:System}(a), containing a capacitor with capacitance $C=0.05$ pF, a typical capacitance for a mammalian neuronal membrane with an area of order $\sim 0.1\;\mu\text{m}^2$ \cite{Gentet2000DirectNeurons,Major1994DetailedSlices}, i.e.\ of the same order as the cross-sectional area of a channel. This capacitor is connected in parallel with four BP conical channels with conductances $g_{+}(t)$, $g_{-}(t)$, $g_{\mathrm{s}}(t)$, and $g_{\mathrm{ss}}(t)$, and four batteries each in series with the conical channels. The channels are taken to be of varying lengths $L_{\pm}=1\text{ }\mu\text{m}$, $L_{\mathrm{s}}=15\text{ }\mu\text{m}$, and $L_{\mathrm{ss}}=90\text{ }\mu\text{m}$. Through Eq.~(\ref{eq:ts}) this translates to timescales $\tau_{\pm}\approx 0.042$ ms for the two fast channels, $\tau_{\mathrm{s}}\approx 9.4 \text{ ms}$ for the slow channel, and $\tau_{\mathrm{ss}}\approx 338 \text{ ms }$ for the super slow channel.  The batteries have potentials $E_{\pm}=\pm 114\text{ mV}$ for the two fast channels, and $E_{\mathrm{s}}=E_{\mathrm{ss}}=-180\text{ mV}$ for the slow and super slow channels. These batteries, which mimic the Nernst potential caused by ionic concentration differences inside and outside the neuron in the HH model \cite{Hodgkin1952ANerve}, are considered to be actual batteries in the microfluidic circuit of interest here, but their potentials are comparable to their biological Nernst potential counterparts \cite{fundNeuroTrain}.

In Fig.~\ref{fig:System}(b) we show a schematic depiction of a BP channel of length $L_i$, implemented in the circuit in Fig.~\ref{fig:System}(a), with base- and tip radii $R_{i,\mathrm{b}}$ and $R_{i,\mathrm{t}}=R_{i,\mathrm{b}}-\Delta R_i$, respectively, and thus with radius $R_i(x)=R_{i,\mathrm{b}}-x\Delta R_i/L_i$ for positions $x\in\left[0,L_i\right]$ in the channel. The channel connects two 1:1 aqueous electrolyte reservoirs with the viscosity $\eta=1.01\text{ mPa}\cdot\text{s}$ and the electric permittivity $\epsilon=0.71\text{ nF}\cdot\text{m}^{-1}$ of water. The cationic and anionic bulk concentrations are given by $\rho_{\mathrm{b}}=2$ mM, comparable to the extracellular potassium concentration in biological neurons \cite{fundNeuroTrain}, which gives rise to a Debye length $\lambda_D\approx 6.8\text{ nm}$. The channels carry a surface charge that linearly decreases from $e\sigma_0=0.1\;e\text{nm}^{-2}$ at the broad base to $-0.05\;e\text{nm}^{-2}$ at the narrow tip, thereby changing by $\sigma^{\prime}=-3\sigma_0/2$ over the channel length and forming a bipolar surface charge profile. These charge densities correspond to Gouy-Chapman zeta potentials that vary between $92\text{ mV}$ and $-61\text{ mV}$. For the short fast channels and the slow channel we fix $R_{i,\mathrm{b}}=200$ nm and $R_{i,\mathrm{t}}=50$ nm, while the super slow channel is narrower with $R_{\mathrm{ss},\mathrm{b}}=120$ nm and $R_{\mathrm{ss},\mathrm{t}}=30$ nm. Thus, in all cases the channel radii are substantially larger than the Debye length, such that overlap of electric double layers is not prominent.  

To fully resolve the dynamics of the circuit depicted in Fig.~\ref{fig:System}(a), we have to know how the conductances $g_i(t)$ of the BP channels evolve. For this we use an analytical model that quantitatively describes the steady-state and dynamical conductance properties of BP channels \cite{Kamsma2023UnveilingIontronics}. BP channels exhibit voltage-dependent salt concentration polarisation in steady-state, with the radially averaged salt concentration $\overline{\rho}_{i,\mathrm{s}}(x,V_i)$ described by
\begin{equation}\label{eq:rhos}
\begin{split}
    \overline{\rho}_{i,\mathrm{s}}(x,V_i)=2\rho_{\mathrm{b}}-\frac{1}{\text{Pe}_i(V_i)/V_i}\frac{2e \left(\sigma_{0}\Delta R_i+\sigma^{\prime} R_{i,\mathrm{b}}\right)}{k_{\mathrm{B}}T  R_{i,\mathrm{t}}^2}\\
    \left(\frac{R_{i,\mathrm{b}}(1-x/L_i)}{R_i(x)}-\frac{e^{-\text{Pe}_i(V_i)\frac{(1-x/L_i)R_{i,\mathrm{t}}}{R_i(x)}}-1}{e^{-\text{Pe}_i(V_i)\frac{R_{i,\mathrm{t}}}{R_{i,\mathrm{b}}}}-1}\right),
    \end{split}
\end{equation}
with $\text{Pe}_i(V_i)=Q_i(V_i)L_i/(\pi DR_{i,\mathrm{t}}^2)$ the P\'{e}clet number at the narrow end and $Q_i(V_i)=-\pi R_{i,\mathrm{t}}R_{i,\mathrm{b}}\epsilon\psi_{\mathrm{eff}} V_i/(\eta L_i)$ the volume flow through the channel. The system is considered to be at a temperature of $293.15$ K and the effective surface potential $\psi_{\mathrm{eff}}=-25$ mV is taken to be the same as in Ref.~\cite{Kamsma2023UnveilingIontronics} as we consider the same surface charge distributions here. The accumulation or depletion of salt affects the conductance of the channel according to
\begin{align}\label{eq:cond}
    g_{i,\infty}(V_i)=&g_{i,0}\frac{L_i}{2\rho_{\mathrm{b}}\int_{0}^{L_i}\left(\overline{\rho}_{i,\mathrm{s}}(x,V_i)\right)^{-1}\dd x},
\end{align}
with $g_{i,0}=(\pi R_{i,\mathrm{t}} R_{i,\mathrm{b}}/L_i)(2\rho_{\rm{b}}e^2D/k_{\mathrm{B}}T)$ the homogeneous channel conductance. In the numerical evaluation of Eq.~(\ref{eq:cond}) we replace $\overline{\rho}_{i,\mathrm{s}}(x,V_i)$ by $\text{Max}\left[0.2
\rho_{\mathrm{b}},\overline{\rho}_{i,\mathrm{s}}(x,V_i)\right]$ to avoid nonphysical negative concentrations that can emerge due to the strong voltage-dependent salt depletion of BP channels \cite{Kamsma2023UnveilingIontronics}. This approach does induce a sharper drop in conductance, compared to full finite-element simulations, when concentrations start to approach the imposed minimum of $0.2\rho_{\mathrm{b}}$, discussed in  more detail in the Supplemental Material. This artefact complicates the circuit equations we introduce below. To help smooth over this sharper drop we employ a third-order interpolation to evaluate Eq.~(\ref{eq:cond}) between voltages spaced at intervals of 0.025 V, ranging from -0.3125 to 0.3125 V. A more sophisticated theoretical model of individual channels in the future should obviate the need for such an ad hoc approach, but for now this effective method suffices.

Since it takes a typical time $\tau_i$ as per Eq.~(\ref{eq:ts}) for salt to accumulate or deplete, the channel exhibits a (volatile) memory conductance with typical memory retention time $\tau_i$. The resulting dynamic conductance $g_i(t)$ was found to be well described by
\begin{align}
    \dfrac{\dd g_i(t)}{\dd t}=\frac{g_{i,\infty}(V_i(t))-g_i(t)}{\tau_i}\label{eq:dgdt},
\end{align}
where $V_i(t)$ is the potential difference between base and tip of the channel, $g_{i,\infty}(V_i)$ is the voltage-dependent steady-state conductance of the channel as per Eq.~(\ref{eq:cond}), and $\tau_i$ is the typical conductance memory retention timescale of the channel given by Eq.~(\ref{eq:ts}) \cite{Kamsma2023UnveilingIontronics}.

With differential equations for each of the dynamic conductances $g_i(t)$, we only need one additional equation to close the set that describes the time-evolution of the ``membrane'' potential $V_{\mathrm{m}}(t)$, here the potential over the capacitor. This additional equation is provided by Kirchhoff's law
\begin{align}
	C\dfrac{\dd V_{\mathrm{m}}(t)}{\dd t}=I(t)-\sum_{i}g_{i}(t)\left(V_{\mathrm{m}}(t)-E_{i}\right),\label{eq:CircuitEq}
\end{align}
where $i\in\left\{+,-,\mathrm{s},\mathrm{ss}\right\}$ and the conductances $g_i(t)$ each evolve according to Eq.~(\ref{eq:dgdt}) with their corresponding $g_{i,\infty}(V_i(t))$ and $\tau_i$. The voltage arguments $V_i(t)$ over the channels are given by $V_-(t)=V_{\mathrm{m}}(t)-E_-$, $V_+(t)=-V_{\mathrm{m}}(t)+E_+$, $V_{\mathrm{s}}(t)=-V_{\mathrm{m}}(t)+E_{\mathrm{s}}$, and $V_{\mathrm{ss}}(t)=-V_{\mathrm{m}}(t)+E_{\mathrm{ss}}$, with the different signs of the potentials corresponding to the different orientations of the channels as depicted in Fig.~\ref{fig:System}(a). Using the initial conditions $V(0)=-70\text{ mV}$ and $g_{i}(0)=g_{i,0}$, with $g_{i,0}$ as defined below Eq.~(\ref{eq:cond}), we numerically solve the closed set of Eqs.~(\ref{eq:ts}), (\ref{eq:dgdt}) and (\ref{eq:CircuitEq}) for various current stimuli $I(t)$. The system is given at least 10 s to settle into a steady state before applying a current $I(t)$, we offset the time in the results to omit this in the plots.

Note that $\tau_{\pm}\ll\tau_{\mathrm{s}},\tau_{\mathrm{ss}}$, additionally $\tau_{\pm}$ is much faster than the typical response time of $V_{\mathrm{m}}(t)$ too as we will see later, so the two short channels actually act as quasi-instantaneous current rectifiers due to their comparatively fast dynamics, rather than memristors, as we will more extensively discuss in Sec.~\ref{sec:channelrole}.

\begin{figure*}[ht]
		\centering
		\includegraphics[width=1\textwidth]{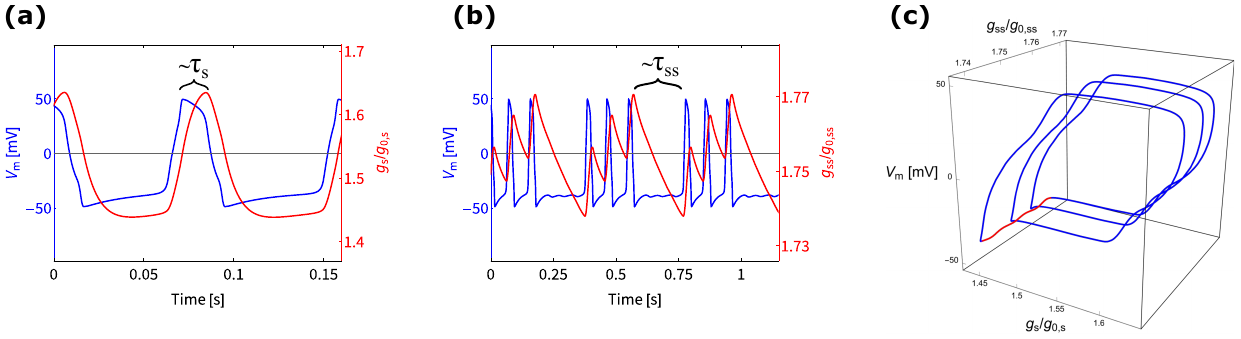}
		\caption{Evolution of the three (relevant) dynamic variables $V_{\mathrm{m}}(t)$, $g_{\mathrm{s}}(t)$, and $g_{\mathrm{ss}}(t)$ during the limit cycle of the tonic bursting case shown in Fig.~\ref{fig:SpikeGrid}(e). \textbf{(a)} Dynamic variables $V_{\mathrm{m}}(t)$ and $g_{\mathrm{s}}(t)$ during a single spike. At low $g_{\mathrm{s}}(t)$ the potential $V_{\mathrm{m}}(t)$ can switch to a positive voltage. Due to the delayed increase of $g_{\mathrm{s}}(t)$ after a timescale $\tau_{\mathrm{s}}$ in response to the increase of $V_{\mathrm{m}}(t)$, the conductance $g_{\mathrm{s}}(t)$ drives $V_{\mathrm{m}}(t)$ down again, forming the spike. \textbf{(b)} Dynamic variables $V_{\mathrm{m}}(t)$ and $g_{\mathrm{ss}}(t)$ during a burst of spikes. After three spikes $g_{\mathrm{ss}}(t)$ increased sufficiently, in response to the increase of $V_{\mathrm{m}}(t)$ during the spikes, to halt the spiking. Without spiking, $g_{\mathrm{ss}}(t)$ decreases over a timescale $\tau_{\mathrm{ss}}$ after which the spiking starts again, creating tonic bursting. \textbf{(c)} The three dynamic variables $V_{\mathrm{m}}(t)$, $g_{\mathrm{s}}(t)$, and $g_{\mathrm{ss}}(t)$ in a phase portrait during tonic bursting, showing a characteristic bursting phase diagram with three loops (blue) where $g_{\mathrm{ss}}(t)$ changes little, corresponding to the three spikes, and a trajectory connecting the third and first loop (red) where $g_{\mathrm{ss}}(t)$ returns from high to low, corresponding to the periods of quiescence.}
		\label{fig:dynamicVariables}
\end{figure*}

\section{Advanced iontronic spiking modes}
Upon numerically evaluating the membrane potential $V_{\mathrm{m}}(t)$ that emerges from the proposed fluidic iontronic circuit introduced in Sec.~\ref{sec:circuit} for various stimuli, we reveal the remarkable diversity of typical neuronal firing modes \cite{Izhikevich2004WhichNeurons} shown in Fig.~\ref{fig:SpikeGrid}, which we will discuss individually below. We stress that all spiking modes discussed below originate from one and the same iontronic circuit, with the stimulus current $I(t)$ the only difference between the spiking modes. Additionally, we note that all spikes exhibit voltage amplitudes and spiking frequencies that are typical for mammalian neurons \cite{Bean2007TheNeurons}.

\subsection{Tonic Spiking and Phasic Spiking}\label{sec:TSPS}
The earlier reported foundational tonic \cite{Robin2021ModelingSlits,Kamsma2023IontronicMemristors} and phasic spiking \cite{Kamsma2023IontronicMemristors} also emerge from the circuit we consider here. Tonic spiking, characterized by a regular train of voltage spikes as shown in Fig.~\ref{fig:SpikeGrid}(a), and phasic spiking, featuring a single isolated voltage spike as shown in Fig.~\ref{fig:SpikeGrid}(b), appear for the present system parameters under sustained current stimuli of 18.40 pA and 19.05 pA, respectively. The phasic spiking current stimulus of 18.40 pA is just above the threshold for any spiking to occur, unless we consider the variability of the threshold as discussed in Sec.~\ref{sec:TV}. The sustained current of Fig.~\ref{fig:SpikeGrid}(b) does give rise, after the single voltage pulse, to a steady voltage that differs from the initial voltage. An all-or-none spike can also appear upon a pulse stimulus, after which the voltage settles back to its initial steady-state \cite{Kamsma2023UnveilingIontronics}. 

The dynamics here are governed by the typical RC-like time of the circuit that determines the time it takes for the (de)polarisation of $V_{\mathrm{m}}(t)$, while the timescale $\tau_{\mathrm{s}}$ dictates the typical width of a spike; the short channels respond on such fast timescales that their dynamics can be assumed to be instantaneous \cite{Kamsma2023IontronicMemristors,kamsma2023math}. Although the timescale $\tau_{\mathrm{ss}}$ does not play a role in these spiking modes as these also appear without the super slow channel \cite{Kamsma2023UnveilingIontronics}, a small influence of the super slow channel is still visible in the case of tonic spiking. The spiking frequency initially is slightly higher immediately after the stimulus is applied and then gradually settles into a lower frequency over a time $\sim\tau_{\mathrm{ss}}$. This actually corresponds to the spiking mode of \textit{spike frequency adaptation} \cite{Izhikevich2004WhichNeurons}, but since this effect is so minor in our results, we choose not to explicitly distinguish it as an additional emerging spiking mode.

\subsection{Phasic Bursting}
Imposing a sustained current stimulus to the circuit of 19.01 pA elicits phasic bursting, a spiking mode where a burst of spikes occurs, followed by a return to a (new) steady state, despite the sustained stimulus. This mode is made possible by the super slow channel. The initial burst has a duration of the typical timescale $\sim\tau_{\mathrm{ss}}$ of the super slow channel, after which this channel has had sufficient time to increase its conductance to return the system to a steady state.

\subsection{Mixed Mode}
Under a sustained stimulus of 19.02 pA we find mixed mode spiking, i.e.\ the iontronic circuit transitions from an initial high-frequency burst of spikes with a duration of $\sim\tau_{\mathrm{ss}}$, into a lower frequency tonic spiking, with the individual spikes now separated by $\sim\tau_{\mathrm{ss}}$, as shown in Fig.~\ref{fig:SpikeGrid}(d). In this case the initial burst is a transient, of typical time $\sim\tau_{\mathrm{ss}}$, as the system settles into the periodic solution of the tonic spiking.

\subsection{Tonic Bursting}
Tonic bursting entails short bursts of spiking interspersed with periods of quiescence. When imposing a sustained stimulus of 19.04 pA we find that the circuit exhibits a periodic behaviour of high frequency burst as shown in Fig.~\ref{fig:SpikeGrid}(e). The durations of the bursts and periods of quiescence are dictated by the slow dynamics of longest channel, as the super slow channel periodically increases and decreases in conductance, visible by the fact that each burst or quiescence period has a duration of order $\sim\tau_{\mathrm{ss}}\approx 338$ ms.

\subsection{Threshold Variability}\label{sec:TV}
Our findings also unveil threshold variability, wherein the firing threshold of the neuron is influenced by prior activity. As shown in Fig.~\ref{fig:SpikeGrid}(f), when imposing a negative and positive stimulus pulse of magnitude $\pm 18.30$ pA (just below the threshold mentioned in Sec.~\ref{sec:TSPS}) of duration 0.02 s, separated by 0.18 s (between the end of the first pulse and the beginning of the second) no spike occurs. However, when we impose precisely the same pulses but now separated by 0.01 s we find that a full spike occurs. Thus in the first set of pulses, the threshold for spiking was not reached, but in the second instance it was reached with exactly the same pulses, showing that the prior activity of the circuit can influence the threshold for spiking. This is a result of the slow channel with timescale $\tau_{\mathrm{s}}$ decreasing in conductance as a result of the negative pulse, while the super slow channel actually plays no role in this spiking mode as it is also observed without the super slow channel. If the interval is much larger than $\tau_{\mathrm{s}}\approx 9.4$ ms, as it is for the first set of pulses, then the slow channel reverts to its steady-state before the second pulse. However, if the interval between the stimuli is of the order of $\tau_{\mathrm{s}}=9.4$ ms (or smaller), as is the case in the second set of pulses where the interval is 10 ms, then the slow channel still has a lowered conductance when the second pulse arrives, making the system more susceptible to stimuli and thereby lowering the firing threshold. 

\subsection{Roles of the different channels}\label{sec:channelrole}
To elucidate the circuit design as shown in Fig.~\ref{fig:System}(b) we heuristically describe the roles the various channels play. Firstly, since $\tau_{\pm}$ is much shorter than the typical response time of $V_{\mathrm{m}}(t)$ and since $\tau_{\pm}\ll\tau_{\mathrm{s}},\tau_{\mathrm{ss}}$, the two short channels actually act as quasi-instantaneous current rectifiers, rather than memristors, though we solve the dynamic equation for all channels for completeness. Tonic and phasic spiking, which occur without the super slow channel, were already remarked to also emerge by using the instantaneous conductance $g_{\infty,\pm}(V(t))$ \cite{Kamsma2023IontronicMemristors} in Eq.~(\ref{eq:CircuitEq}), reducing such a three channel circuit to a two-dimensional dynamic system with dynamic variables $V_{\mathrm{m}}(t)$ and $g_{\mathrm{s}}(t)$ \cite{kamsma2023math}. By extension the results we present here are represented by the three dynamic variables $V_{\mathrm{m}}(t)$, $g_{\mathrm{s}}(t)$, and $g_{\mathrm{ss}}(t)$. In Fig.~\ref{fig:dynamicVariables} we show these dynamic variables during the tonic bursting shown in Fig.~\ref{fig:SpikeGrid}(e).

All channels drive $V_{\mathrm{m}}(t)$ toward their respective battery potentials. When $V_{\mathrm{m}}(t)$ is near $E_{\pm}$ the respective corresponding short fast channel has a high conductance and maintains $V_{\mathrm{m}}(t)$ in that state, forming temporary stable states between which $V_{\mathrm{m}}(t)$ switches during spiking. The slow channel, which drives $V_{\mathrm{m}}(t)$ towards $E_{\mathrm{s}}=-180$ mV, is in a low conductance state when $V_{\mathrm{m}}(t)$ is negative, allowing $V_{\mathrm{m}}(t)$ to depolarize and switch to the positive voltage state upon a stimulus. Following the increase in $V_{\mathrm{m}}(t)$, the slow channel increases in conductance over a timescale $\tau_{\mathrm{s}}$ as we show in Fig.~\ref{fig:dynamicVariables}(a), resulting in a consequent downward shift of $V_{\mathrm{m}}(t)$. This behavior analogously resembles the delayed activation of $\text{K}^{+}$ channels in the Hodgkin-Huxley model \cite{Hodgkin1952ANerve}. The super slow channel, operating over a timescale $\tau_{\mathrm{ss}}$, plays a role akin to the slow channel. However, as we show in Fig.~\ref{fig:dynamicVariables}(b), it takes several spikes for the super slow channel to sufficiently increase in conductance and drive $V_{\mathrm{m}}(t)$ towards its negative battery potential $E_{\mathrm{ss}}=-180$ mV, consequently suppressing spiking. The resulting period of quiescence lasts a time $\sim\tau_{\mathrm{ss}}$, forming a bursting process. The salt accumulation and depletion underpinning the conductance change of the super slow channel bear similarity to the slow intracellular $\text{Ca}^{2+}$ accumulation and depletion implicated in regulating bursting in biological neurons \cite{Chay1983EyringOscillations,Chay1985ChaosCell,Xu2020BifurcationsCircuit}. Combining the three relevant dynamic variables $V_{\mathrm{m}}(t)$, $g_{\mathrm{s}}(t)$, and $g_{\mathrm{ss}}(t)$ in one phase portrait yields Fig.~\ref{fig:dynamicVariables}(c), revealing a characteristic bursting trajectory with three loops (blue) and a path connecting the third and first loop (red), corresponding to the three spikes and to the periods of quiescence, respectively.

\section{Discussion and conclusion}
Previously reported fluidic iontronic circuits have demonstrated tonic spiking \cite{Robin2021ModelingSlits,Kamsma2023IontronicMemristors} and phasic spiking \cite{Kamsma2023IontronicMemristors}. In this study, we extend the repertoire of emergent spiking modes by introducing a new HH-like fluidic iontronic circuit, consisting of a capacitor and four iontronic memristors, that exhibits phasic bursting, mixed-mode spiking, tonic bursting, and threshold variability \cite{Izhikevich2004WhichNeurons}, as well as the earlier reported tonic and phasic spiking \cite{Robin2021ModelingSlits,Kamsma2023IontronicMemristors}. The spikes in our proposed modes exhibit voltages and frequencies that align with those observed in mammalian neurons \cite{Bean2007TheNeurons}. Moreover, the capacitance, battery potentials and salt concentration in the circuit are comparable to their biological counterparts \cite{fundNeuroTrain}. Our theoretical framework builds upon a previously proposed iontronic circuit that exhibits tonic and phasic spiking \cite{Kamsma2023IontronicMemristors} and a physical model for conical ion channels with a bipolar (rather than unipolar) surface charge \cite{Kamsma2023UnveilingIontronics}. These channels are memristive \cite{Kamsma2023UnveilingIontronics} and their typical conductance memory retention time is dependent on the channel length. By varying the lengths of the four channels we can incorporate timescales on the order of a single spike and of entire bursts in a single circuit, allowing for the spiking and bursting processes that emerge from one and the same circuit.

While our theoretical framework in principle is fully physical, a limitation is the parameter sensitivity of the system, at least for the system parameters we considered. The stimuli strengths that induce different spiking modes are only separated by $\sim 0.01-0.1$ pA on the scale of about 20 pA. Notably, if wider current stimuli intervals are found for spiking, then it is possible that \textit{class 2 spiking} \cite{Izhikevich2004WhichNeurons} can also be distinguished as a separate feature as the transitions in frequency seem to be discontinuous, but class 1 or 2 spiking is typically evaluated over varying stimuli intervals which in our case are too narrow to meaningfully investigate this. Additionally, although spiking was found to emerge for a wide range of different parameter configurations, the spiking is sensitive to small individual changes of the short fast channels or their respective batteries, where (at least one mode of) spiking only appeared in the tight interval $E_{\pm}\in\pm\left[113,114.8\right]$ mV. The short fast channels play no dynamic roles in the circuit, but rather act as instantaneous current rectifiers that create the stable voltage states between which $V_{\mathrm{m}}$ oscillates during spiking due to the dynamic switching of the (super) slow channel. Hence, no memristive properties are required for the short fast channels, which only offer a current rectification of around $\approx 21$ \cite{Kamsma2023UnveilingIontronics}, and other (perhaps better performing) diodic devices from the wide range of iontronic current rectifiers \cite{RizaPutra2021MicroscaleOverview,Cheng2010NanofluidicDiodes,Lan2016Voltage-RectifiedNanopores,Kim2022AsymmetricDetection,Choi2016HighMembrane} could be considered. Although some devices can be described by similar theoretical models as we use here \cite{Choi2016HighMembrane,Kamsma2023Brain-inspiredNanochannels}, our analysis is limited to devices for which we have an analytical quantitative model, but iontronic devices that lack such models are still feasible for experimental fabrication. For the (super) slow channels we do require the specific (length-dependent) volatile dynamics we find for the iontronic memristors of concern here, but in this case the system is far more stable against parameter shifts, thereby supporting the use of the (super) slow channels as described here. At least one mode of spiking emerges for changing a single parameter at a time in the range $E_{\mathrm{s}}\in\left[-200,-90\right]$ mV, $E_{\mathrm{ss}}\in\left[-450,-140\right]$ mV, while the capacitance can even span orders of magnitude $C\in[10^{-6},10^{-1}]$ pF. Therefore, the three components that govern the circuit dynamics, i.e.\ the (super) slow channels and the capacitor, are relatively robust once the short fast channels are in order.

The above suggestion for future improvements using other fluidic devices is supported by the fact that the results presented here are already an expansion on results we derived earlier for simpler unipolar conical channels carrying a homogeneous surface charge. Tonic bursting also emerges from a similar circuit with unipolar channels, but with circuit parameters (i.e.\ higher battery potentials, lower salt concentration, lower capacitance) and spiking voltages that are further removed from their biological analogs. The results and specific parameters for the unipolar channel circuit are laid out in the Supplemental Material. The emergence of tonic bursting in a different circuit with different fluidic memristors shows that the bursting spiking modes we present are not inherently dependent on the bipolar conical channels we consider here. Therefore, possible further improvements can be achieved by considering fluidic iontronic devices with an even wider range of attainable conductances. However, this is an issue of individual device physics and here we mostly focused on the overall circuit architecture and the spiking modes it enables.

In summary, we have considerably expanded the range of spiking modes proposed to emerge from iontronic fluidic circuits, entailing phasic bursting, mixed-mode spiking, tonic bursting, and threshold variability. The alignment of the spikes in our results with typical mammalian neuronal voltages and frequencies, combined with various circuit parameters that are comparable to their biological counterparts, further supports the potential that fluidic iontronics carry for neuromorphic spiking circuits. Moreover, since these biologically realistic spikes emerge from a circuit that is based upon the same aqueous electrolyte medium as in neurons, a unique perspective is the future possible integration with biological systems. However, the present system is rather sensitive to stimulus strengths and other circuit parameters, especially in the short fast channels, a limitation that may be mitigated by implementing fluidic devices with a broader range of available conductances. Nevertheless, we showed that the multiscale diffusive timescales of fluidic iontronic memristors of different lengths facilitate a relatively simple circuit that exhibits various advanced modes of neuronal spiking. Consequently, this work contributes to the ongoing exploration of fluidic iontronics as a promising platform for neuromorphic circuits, providing theoretical insights and proposed applications, thereby paving the way for future advancements in this burgeoning field.

\begin{acknowledgments}
This work is part of the D-ITP consortium, a program of the Netherlands Organisation for Scientific Research (NWO) that is funded by the Dutch Ministry of Education, Culture and Science (OCW).
\end{acknowledgments}


\begin{thebibliography}{61}%
\makeatletter
\providecommand \@ifxundefined [1]{%
 \@ifx{#1\undefined}
}%
\providecommand \@ifnum [1]{%
 \ifnum #1\expandafter \@firstoftwo
 \else \expandafter \@secondoftwo
 \fi
}%
\providecommand \@ifx [1]{%
 \ifx #1\expandafter \@firstoftwo
 \else \expandafter \@secondoftwo
 \fi
}%
\providecommand \natexlab [1]{#1}%
\providecommand \enquote  [1]{``#1''}%
\providecommand \bibnamefont  [1]{#1}%
\providecommand \bibfnamefont [1]{#1}%
\providecommand \citenamefont [1]{#1}%
\providecommand \href@noop [0]{\@secondoftwo}%
\providecommand \href [0]{\begingroup \@sanitize@url \@href}%
\providecommand \@href[1]{\@@startlink{#1}\@@href}%
\providecommand \@@href[1]{\endgroup#1\@@endlink}%
\providecommand \@sanitize@url [0]{\catcode `\\12\catcode `\$12\catcode `\&12\catcode `\#12\catcode `\^12\catcode `\_12\catcode `\%12\relax}%
\providecommand \@@startlink[1]{}%
\providecommand \@@endlink[0]{}%
\providecommand \url  [0]{\begingroup\@sanitize@url \@url }%
\providecommand \@url [1]{\endgroup\@href {#1}{\urlprefix }}%
\providecommand \urlprefix  [0]{URL }%
\providecommand \Eprint [0]{\href }%
\providecommand \doibase [0]{http://dx.doi.org/}%
\providecommand \selectlanguage [0]{\@gobble}%
\providecommand \bibinfo  [0]{\@secondoftwo}%
\providecommand \bibfield  [0]{\@secondoftwo}%
\providecommand \translation [1]{[#1]}%
\providecommand \BibitemOpen [0]{}%
\providecommand \bibitemStop [0]{}%
\providecommand \bibitemNoStop [0]{.\EOS\space}%
\providecommand \EOS [0]{\spacefactor3000\relax}%
\providecommand \BibitemShut  [1]{\csname bibitem#1\endcsname}%
\let\auto@bib@innerbib\@empty
\bibitem [{\citenamefont {Schuman}\ \emph {et~al.}(2022)\citenamefont {Schuman}, \citenamefont {Kulkarni}, \citenamefont {Parsa}, \citenamefont {Mitchell}, \citenamefont {Date},\ and\ \citenamefont {Kay}}]{Schuman2022OpportunitiesApplications}%
  \BibitemOpen
  \bibfield  {author} {\bibinfo {author} {\bibfnamefont {C.~D.}\ \bibnamefont {Schuman}}, \bibinfo {author} {\bibfnamefont {S.~R.}\ \bibnamefont {Kulkarni}}, \bibinfo {author} {\bibfnamefont {M.}~\bibnamefont {Parsa}}, \bibinfo {author} {\bibfnamefont {J.~P.}\ \bibnamefont {Mitchell}}, \bibinfo {author} {\bibfnamefont {P.}~\bibnamefont {Date}}, \ and\ \bibinfo {author} {\bibfnamefont {B.}~\bibnamefont {Kay}},\ }\bibfield  {title} {\enquote {\bibinfo {title} {{Opportunities for neuromorphic computing algorithms and applications}},}\ }\href {\doibase 10.1038/s43588-021-00184-y} {\bibfield  {journal} {\bibinfo  {journal} {Nature Computational Science}\ }\textbf {\bibinfo {volume} {2}},\ \bibinfo {pages} {10--19} (\bibinfo {year} {2022})}\BibitemShut {NoStop}%
\bibitem [{\citenamefont {Sangwan}\ and\ \citenamefont {Hersam}(2020)}]{Sangwan2020NeuromorphicMaterials}%
  \BibitemOpen
  \bibfield  {author} {\bibinfo {author} {\bibfnamefont {V.~K.}\ \bibnamefont {Sangwan}}\ and\ \bibinfo {author} {\bibfnamefont {M.~C.}\ \bibnamefont {Hersam}},\ }\bibfield  {title} {\enquote {\bibinfo {title} {{Neuromorphic nanoelectronic materials}},}\ }\href {\doibase 10.1038/s41565-020-0647-z} {\bibfield  {journal} {\bibinfo  {journal} {Nature Nanotechnology}\ }\textbf {\bibinfo {volume} {15}},\ \bibinfo {pages} {517--528} (\bibinfo {year} {2020})}\BibitemShut {NoStop}%
\bibitem [{\citenamefont {Schuman}\ \emph {et~al.}(2017)\citenamefont {Schuman}, \citenamefont {Potok}, \citenamefont {Patton}, \citenamefont {Birdwell}, \citenamefont {Dean}, \citenamefont {Rose},\ and\ \citenamefont {Plank}}]{Schuman2017AHardware}%
  \BibitemOpen
  \bibfield  {author} {\bibinfo {author} {\bibfnamefont {C.~D.}\ \bibnamefont {Schuman}}, \bibinfo {author} {\bibfnamefont {T.~E.}\ \bibnamefont {Potok}}, \bibinfo {author} {\bibfnamefont {R.~M.}\ \bibnamefont {Patton}}, \bibinfo {author} {\bibfnamefont {J.~D.}\ \bibnamefont {Birdwell}}, \bibinfo {author} {\bibfnamefont {M.~E.}\ \bibnamefont {Dean}}, \bibinfo {author} {\bibfnamefont {G.~S.}\ \bibnamefont {Rose}}, \ and\ \bibinfo {author} {\bibfnamefont {J.~S.}\ \bibnamefont {Plank}},\ }\bibfield  {title} {\enquote {\bibinfo {title} {{A Survey of Neuromorphic Computing and Neural Networks in Hardware}},}\ }\href {http://arxiv.org/abs/1705.06963} {\bibfield  {journal} {\bibinfo  {journal} {arXiv}\ } (\bibinfo {year} {2017})}\BibitemShut {NoStop}%
\bibitem [{\citenamefont {Zhu}\ \emph {et~al.}(2020)\citenamefont {Zhu}, \citenamefont {Zhang}, \citenamefont {Yang},\ and\ \citenamefont {Huang}}]{Zhu2020ADevices}%
  \BibitemOpen
  \bibfield  {author} {\bibinfo {author} {\bibfnamefont {J.}~\bibnamefont {Zhu}}, \bibinfo {author} {\bibfnamefont {T.}~\bibnamefont {Zhang}}, \bibinfo {author} {\bibfnamefont {Y.}~\bibnamefont {Yang}}, \ and\ \bibinfo {author} {\bibfnamefont {R.}~\bibnamefont {Huang}},\ }\bibfield  {title} {\enquote {\bibinfo {title} {{A comprehensive review on emerging artificial neuromorphic devices}},}\ }\href {\doibase 10.1063/1.5118217} {\bibfield  {journal} {\bibinfo  {journal} {Applied Physics Reviews}\ }\textbf {\bibinfo {volume} {7}},\ \bibinfo {pages} {011312} (\bibinfo {year} {2020})}\BibitemShut {NoStop}%
\bibitem [{\citenamefont {Sah}\ \emph {et~al.}(2014)\citenamefont {Sah}, \citenamefont {Kim},\ and\ \citenamefont {Chua}}]{Sah2014BrainsMemristors}%
  \BibitemOpen
  \bibfield  {author} {\bibinfo {author} {\bibfnamefont {M.~P.}\ \bibnamefont {Sah}}, \bibinfo {author} {\bibfnamefont {H.}~\bibnamefont {Kim}}, \ and\ \bibinfo {author} {\bibfnamefont {L.~O.}\ \bibnamefont {Chua}},\ }\bibfield  {title} {\enquote {\bibinfo {title} {{Brains are made of memristors}},}\ }\href {\doibase 10.1109/MCAS.2013.2296414} {\bibfield  {journal} {\bibinfo  {journal} {IEEE Circuits and Systems Magazine}\ }\textbf {\bibinfo {volume} {14}},\ \bibinfo {pages} {12--36} (\bibinfo {year} {2014})}\BibitemShut {NoStop}%
\bibitem [{\citenamefont {Hodgkin}\ and\ \citenamefont {Huxley}(1952)}]{Hodgkin1952ANerve}%
  \BibitemOpen
  \bibfield  {author} {\bibinfo {author} {\bibfnamefont {A.~L.}\ \bibnamefont {Hodgkin}}\ and\ \bibinfo {author} {\bibfnamefont {A.~F.}\ \bibnamefont {Huxley}},\ }\bibfield  {title} {\enquote {\bibinfo {title} {{A quantitative description of membrane current and its application to conduction and excitation in nerve}},}\ }\href {\doibase 10.1113/JPHYSIOL.1952.SP004764} {\bibfield  {journal} {\bibinfo  {journal} {The Journal of Physiology}\ }\textbf {\bibinfo {volume} {117}},\ \bibinfo {pages} {500} (\bibinfo {year} {1952})}\BibitemShut {NoStop}%
\bibitem [{\citenamefont {Chua}\ \emph {et~al.}(2012)\citenamefont {Chua}, \citenamefont {Sbitnev},\ and\ \citenamefont {Kim}}]{Chua2012Hodgkin-HuxleyMemristors}%
  \BibitemOpen
  \bibfield  {author} {\bibinfo {author} {\bibfnamefont {L.}~\bibnamefont {Chua}}, \bibinfo {author} {\bibfnamefont {V.}~\bibnamefont {Sbitnev}}, \ and\ \bibinfo {author} {\bibfnamefont {H.}~\bibnamefont {Kim}},\ }\bibfield  {title} {\enquote {\bibinfo {title} {{Hodgkin-Huxley axon is made of memristors}},}\ }\href {\doibase 10.1142/S021812741230011X} {\bibfield  {journal} {\bibinfo  {journal} {https://doi-org.proxy.library.uu.nl/10.1142/S021812741230011X}\ }\textbf {\bibinfo {volume} {22}},\ \bibinfo {pages} {1230011} (\bibinfo {year} {2012})}\BibitemShut {NoStop}%
\bibitem [{\citenamefont {Caravelli}\ and\ \citenamefont {Carbajal}(2018)}]{Caravelli2018MemristorsOutsiders}%
  \BibitemOpen
  \bibfield  {author} {\bibinfo {author} {\bibfnamefont {F.}~\bibnamefont {Caravelli}}\ and\ \bibinfo {author} {\bibfnamefont {J.~P.}\ \bibnamefont {Carbajal}},\ }\bibfield  {title} {\enquote {\bibinfo {title} {{Memristors for the Curious Outsiders}},}\ }\href {\doibase 10.3390/TECHNOLOGIES6040118} {\bibfield  {journal} {\bibinfo  {journal} {Technologies 2018, Vol. 6, Page 118}\ }\textbf {\bibinfo {volume} {6}},\ \bibinfo {pages} {118} (\bibinfo {year} {2018})}\BibitemShut {NoStop}%
\bibitem [{\citenamefont {Chua}(2013)}]{Chua2013MemristorChaos}%
  \BibitemOpen
  \bibfield  {author} {\bibinfo {author} {\bibfnamefont {L.}~\bibnamefont {Chua}},\ }\bibfield  {title} {\enquote {\bibinfo {title} {{Memristor, Hodgkin-Huxley, and edge of chaos}},}\ }\href {\doibase 10.1088/0957-4484/24/38/383001} {\bibfield  {journal} {\bibinfo  {journal} {Nanotechnology}\ }\textbf {\bibinfo {volume} {24}} (\bibinfo {year} {2013}),\ 10.1088/0957-4484/24/38/383001}\BibitemShut {NoStop}%
\bibitem [{\citenamefont {Thakur}\ \emph {et~al.}(2018)\citenamefont {Thakur}, \citenamefont {Molin}, \citenamefont {Cauwenberghs}, \citenamefont {Indiveri}, \citenamefont {Kumar}, \citenamefont {Qiao}, \citenamefont {Schemmel}, \citenamefont {Wang}, \citenamefont {Chicca}, \citenamefont {Olson~Hasler}, \citenamefont {Seo}, \citenamefont {Yu}, \citenamefont {Cao}, \citenamefont {van Schaik},\ and\ \citenamefont {Etienne-Cummings}}]{Thakur2018Large-ScaleBrain}%
  \BibitemOpen
  \bibfield  {author} {\bibinfo {author} {\bibfnamefont {C.~S.}\ \bibnamefont {Thakur}}, \bibinfo {author} {\bibfnamefont {J.~L.}\ \bibnamefont {Molin}}, \bibinfo {author} {\bibfnamefont {G.}~\bibnamefont {Cauwenberghs}}, \bibinfo {author} {\bibfnamefont {G.}~\bibnamefont {Indiveri}}, \bibinfo {author} {\bibfnamefont {K.}~\bibnamefont {Kumar}}, \bibinfo {author} {\bibfnamefont {N.}~\bibnamefont {Qiao}}, \bibinfo {author} {\bibfnamefont {J.}~\bibnamefont {Schemmel}}, \bibinfo {author} {\bibfnamefont {R.}~\bibnamefont {Wang}}, \bibinfo {author} {\bibfnamefont {E.}~\bibnamefont {Chicca}}, \bibinfo {author} {\bibfnamefont {J.}~\bibnamefont {Olson~Hasler}}, \bibinfo {author} {\bibfnamefont {J.~S.}\ \bibnamefont {Seo}}, \bibinfo {author} {\bibfnamefont {S.}~\bibnamefont {Yu}}, \bibinfo {author} {\bibfnamefont {Y.}~\bibnamefont {Cao}}, \bibinfo {author} {\bibfnamefont {A.}~\bibnamefont {van Schaik}}, \ and\ \bibinfo {author} {\bibfnamefont {R.}~\bibnamefont {Etienne-Cummings}},\ }\bibfield  {title} {\enquote
  {\bibinfo {title} {{Large-Scale Neuromorphic Spiking Array Processors: A Quest to Mimic the Brain}},}\ }\href {\doibase 10.3389/FNINS.2018.00891/BIBTEX} {\bibfield  {journal} {\bibinfo  {journal} {Frontiers in Neuroscience}\ }\textbf {\bibinfo {volume} {12}},\ \bibinfo {pages} {353526} (\bibinfo {year} {2018})}\BibitemShut {NoStop}%
\bibitem [{\citenamefont {Yang}\ \emph {et~al.}(2020)\citenamefont {Yang}, \citenamefont {Wang}, \citenamefont {Ren}, \citenamefont {Mao}, \citenamefont {Wang}, \citenamefont {Zhou}, \citenamefont {Han}, \citenamefont {Yang}, \citenamefont {Wang}, \citenamefont {Han}, \citenamefont {Ren}, \citenamefont {Mao}, \citenamefont {Wang},\ and\ \citenamefont {Zhou}}]{Yang2020NeuromorphicSystems}%
  \BibitemOpen
  \bibfield  {author} {\bibinfo {author} {\bibfnamefont {J.-Q.}\ \bibnamefont {Yang}}, \bibinfo {author} {\bibfnamefont {R.}~\bibnamefont {Wang}}, \bibinfo {author} {\bibfnamefont {Y.}~\bibnamefont {Ren}}, \bibinfo {author} {\bibfnamefont {J.-Y.}\ \bibnamefont {Mao}}, \bibinfo {author} {\bibfnamefont {Z.-P.}\ \bibnamefont {Wang}}, \bibinfo {author} {\bibfnamefont {Y.}~\bibnamefont {Zhou}}, \bibinfo {author} {\bibfnamefont {S.-T.}\ \bibnamefont {Han}}, \bibinfo {author} {\bibfnamefont {J.-q.}\ \bibnamefont {Yang}}, \bibinfo {author} {\bibfnamefont {R.}~\bibnamefont {Wang}}, \bibinfo {author} {\bibfnamefont {S.-t.}\ \bibnamefont {Han}}, \bibinfo {author} {\bibfnamefont {Y.}~\bibnamefont {Ren}}, \bibinfo {author} {\bibfnamefont {J.-y.}\ \bibnamefont {Mao}}, \bibinfo {author} {\bibfnamefont {Z.-p.}\ \bibnamefont {Wang}}, \ and\ \bibinfo {author} {\bibfnamefont {Y.}~\bibnamefont {Zhou}},\ }\bibfield  {title} {\enquote {\bibinfo {title} {{Neuromorphic Engineering: From Biological to Spike-Based Hardware Nervous
  Systems}},}\ }\href {\doibase 10.1002/ADMA.202003610} {\bibfield  {journal} {\bibinfo  {journal} {Advanced Materials}\ }\textbf {\bibinfo {volume} {32}},\ \bibinfo {pages} {2003610} (\bibinfo {year} {2020})}\BibitemShut {NoStop}%
\bibitem [{\citenamefont {Izhikevich}(2004)}]{Izhikevich2004WhichNeurons}%
  \BibitemOpen
  \bibfield  {author} {\bibinfo {author} {\bibfnamefont {E.~M.}\ \bibnamefont {Izhikevich}},\ }\bibfield  {title} {\enquote {\bibinfo {title} {{Which model to use for cortical spiking neurons?}}}\ }\href {\doibase 10.1109/TNN.2004.832719} {\bibfield  {journal} {\bibinfo  {journal} {IEEE Transactions on Neural Networks}\ }\textbf {\bibinfo {volume} {15}},\ \bibinfo {pages} {1063--1070} (\bibinfo {year} {2004})}\BibitemShut {NoStop}%
\bibitem [{\citenamefont {Bean}(2007)}]{Bean2007TheNeurons}%
  \BibitemOpen
  \bibfield  {author} {\bibinfo {author} {\bibfnamefont {B.~P.}\ \bibnamefont {Bean}},\ }\bibfield  {title} {\enquote {\bibinfo {title} {{The action potential in mammalian central neurons}},}\ }\href {\doibase 10.1038/nrn2148} {\bibfield  {journal} {\bibinfo  {journal} {Nature Reviews Neuroscience 2007 8:6}\ }\textbf {\bibinfo {volume} {8}},\ \bibinfo {pages} {451--465} (\bibinfo {year} {2007})}\BibitemShut {NoStop}%
\bibitem [{\citenamefont {Gray}\ and\ \citenamefont {McCormick}(1996)}]{Gray1996ChatteringCortex}%
  \BibitemOpen
  \bibfield  {author} {\bibinfo {author} {\bibfnamefont {C.~M.}\ \bibnamefont {Gray}}\ and\ \bibinfo {author} {\bibfnamefont {D.~A.}\ \bibnamefont {McCormick}},\ }\bibfield  {title} {\enquote {\bibinfo {title} {{Chattering Cells: Superficial Pyramidal Neurons Contributing to the Generation of Synchronous Oscillations in the Visual Cortex}},}\ }\href {\doibase 10.1126/SCIENCE.274.5284.109} {\bibfield  {journal} {\bibinfo  {journal} {Science}\ }\textbf {\bibinfo {volume} {274}},\ \bibinfo {pages} {109--113} (\bibinfo {year} {1996})}\BibitemShut {NoStop}%
\bibitem [{\citenamefont {Kamsma}\ \emph {et~al.}(2023{\natexlab{a}})\citenamefont {Kamsma}, \citenamefont {Boon}, \citenamefont {Ter~Rele}, \citenamefont {Spitoni},\ and\ \citenamefont {Van~Roij}}]{Kamsma2023IontronicMemristors}%
  \BibitemOpen
  \bibfield  {author} {\bibinfo {author} {\bibfnamefont {T.~M.}\ \bibnamefont {Kamsma}}, \bibinfo {author} {\bibfnamefont {W.~Q.}\ \bibnamefont {Boon}}, \bibinfo {author} {\bibfnamefont {T.}~\bibnamefont {Ter~Rele}}, \bibinfo {author} {\bibfnamefont {C.}~\bibnamefont {Spitoni}}, \ and\ \bibinfo {author} {\bibfnamefont {R.}~\bibnamefont {Van~Roij}},\ }\bibfield  {title} {\enquote {\bibinfo {title} {{Iontronic Neuromorphic Signaling with Conical Microfluidic Memristors}},}\ }\href {\doibase 10.1103/PHYSREVLETT.130.268401/FIGURES/3/MEDIUM} {\bibfield  {journal} {\bibinfo  {journal} {Physical Review Letters}\ }\textbf {\bibinfo {volume} {130}},\ \bibinfo {pages} {268401} (\bibinfo {year} {2023}{\natexlab{a}})}\BibitemShut {NoStop}%
\bibitem [{\citenamefont {Micu}\ \emph {et~al.}(2017)\citenamefont {Micu}, \citenamefont {Plemel}, \citenamefont {Caprariello}, \citenamefont {Nave},\ and\ \citenamefont {Stys}}]{Micu2017Axo-myelinicSystem}%
  \BibitemOpen
  \bibfield  {author} {\bibinfo {author} {\bibfnamefont {I.}~\bibnamefont {Micu}}, \bibinfo {author} {\bibfnamefont {J.~R.}\ \bibnamefont {Plemel}}, \bibinfo {author} {\bibfnamefont {A.~V.}\ \bibnamefont {Caprariello}}, \bibinfo {author} {\bibfnamefont {K.~A.}\ \bibnamefont {Nave}}, \ and\ \bibinfo {author} {\bibfnamefont {P.~K.}\ \bibnamefont {Stys}},\ }\bibfield  {title} {\enquote {\bibinfo {title} {{Axo-myelinic neurotransmission: a novel mode of cell signalling in the central nervous system}},}\ }\href {\doibase 10.1038/NRN.2017.128} {\bibfield  {journal} {\bibinfo  {journal} {Nature Reviews Neuroscience 2017 19:1}\ }\textbf {\bibinfo {volume} {19}},\ \bibinfo {pages} {49--58} (\bibinfo {year} {2017})}\BibitemShut {NoStop}%
\bibitem [{\citenamefont {Pereda}\ and\ \citenamefont {Purpura}(2014)}]{Pereda2014ElectricalSynapses}%
  \BibitemOpen
  \bibfield  {author} {\bibinfo {author} {\bibfnamefont {A.~E.}\ \bibnamefont {Pereda}}\ and\ \bibinfo {author} {\bibfnamefont {D.~P.}\ \bibnamefont {Purpura}},\ }\bibfield  {title} {\enquote {\bibinfo {title} {{Electrical synapses and their functional interactions with chemical synapses}},}\ }\href {\doibase 10.1038/NRN3708} {\bibfield  {journal} {\bibinfo  {journal} {Nature Reviews Neuroscience 2014 15:4}\ }\textbf {\bibinfo {volume} {15}},\ \bibinfo {pages} {250--263} (\bibinfo {year} {2014})}\BibitemShut {NoStop}%
\bibitem [{\citenamefont {Xia}\ and\ \citenamefont {Storm}(2005)}]{Xia2005ThePlasticity}%
  \BibitemOpen
  \bibfield  {author} {\bibinfo {author} {\bibfnamefont {Z.}~\bibnamefont {Xia}}\ and\ \bibinfo {author} {\bibfnamefont {D.~R.}\ \bibnamefont {Storm}},\ }\bibfield  {title} {\enquote {\bibinfo {title} {{The role of calmodulin as a signal integrator for synaptic plasticity}},}\ }\href {\doibase 10.1038/nrn1647} {\bibfield  {journal} {\bibinfo  {journal} {Nature Reviews Neuroscience}\ }\textbf {\bibinfo {volume} {6}},\ \bibinfo {pages} {267--276} (\bibinfo {year} {2005})}\BibitemShut {NoStop}%
\bibitem [{\citenamefont {L{\"{u}}scher}\ and\ \citenamefont {Malenka}(2012)}]{Luscher2012NMDALTP/LTD}%
  \BibitemOpen
  \bibfield  {author} {\bibinfo {author} {\bibfnamefont {C.}~\bibnamefont {L{\"{u}}scher}}\ and\ \bibinfo {author} {\bibfnamefont {R.~C.}\ \bibnamefont {Malenka}},\ }\bibfield  {title} {\enquote {\bibinfo {title} {{NMDA Receptor-Dependent Long-Term Potentiation and Long-Term Depression (LTP/LTD)}},}\ }\href {\doibase 10.1101/CSHPERSPECT.A005710} {\bibfield  {journal} {\bibinfo  {journal} {Cold Spring Harbor Perspectives in Biology}\ }\textbf {\bibinfo {volume} {4}},\ \bibinfo {pages} {1--15} (\bibinfo {year} {2012})}\BibitemShut {NoStop}%
\bibitem [{\citenamefont {Covi}\ \emph {et~al.}(2021)\citenamefont {Covi}, \citenamefont {Donati}, \citenamefont {Liang}, \citenamefont {Kappel}, \citenamefont {Heidari}, \citenamefont {Payvand},\ and\ \citenamefont {Wang}}]{Covi2021AdaptiveDevices}%
  \BibitemOpen
  \bibfield  {author} {\bibinfo {author} {\bibfnamefont {E.}~\bibnamefont {Covi}}, \bibinfo {author} {\bibfnamefont {E.}~\bibnamefont {Donati}}, \bibinfo {author} {\bibfnamefont {X.}~\bibnamefont {Liang}}, \bibinfo {author} {\bibfnamefont {D.}~\bibnamefont {Kappel}}, \bibinfo {author} {\bibfnamefont {H.}~\bibnamefont {Heidari}}, \bibinfo {author} {\bibfnamefont {M.}~\bibnamefont {Payvand}}, \ and\ \bibinfo {author} {\bibfnamefont {W.}~\bibnamefont {Wang}},\ }\bibfield  {title} {\enquote {\bibinfo {title} {{Adaptive Extreme Edge Computing for Wearable Devices}},}\ }\href {\doibase 10.3389/FNINS.2021.611300/BIBTEX} {\bibfield  {journal} {\bibinfo  {journal} {Frontiers in Neuroscience}\ }\textbf {\bibinfo {volume} {15}},\ \bibinfo {pages} {611300} (\bibinfo {year} {2021})}\BibitemShut {NoStop}%
\bibitem [{\citenamefont {Chicca}\ and\ \citenamefont {Indiveri}(2020)}]{Chicca2020ASystems}%
  \BibitemOpen
  \bibfield  {author} {\bibinfo {author} {\bibfnamefont {E.}~\bibnamefont {Chicca}}\ and\ \bibinfo {author} {\bibfnamefont {G.}~\bibnamefont {Indiveri}},\ }\bibfield  {title} {\enquote {\bibinfo {title} {{A recipe for creating ideal hybrid memristive-CMOS neuromorphic processing systems}},}\ }\href {\doibase 10.1063/1.5142089/570949} {\bibfield  {journal} {\bibinfo  {journal} {Applied Physics Letters}\ }\textbf {\bibinfo {volume} {116}},\ \bibinfo {pages} {120501} (\bibinfo {year} {2020})}\BibitemShut {NoStop}%
\bibitem [{\citenamefont {Wang}\ \emph {et~al.}(2022)\citenamefont {Wang}, \citenamefont {Zhang}, \citenamefont {Astier}, \citenamefont {Nickle}, \citenamefont {Soni}, \citenamefont {Alami}, \citenamefont {Borrini}, \citenamefont {Zhang}, \citenamefont {Honnigfort}, \citenamefont {Braunschweig}, \citenamefont {Leoncini}, \citenamefont {Qi}, \citenamefont {Han}, \citenamefont {del Barco}, \citenamefont {Thompson},\ and\ \citenamefont {Nijhuis}}]{Wang2022DynamicBehaviour}%
  \BibitemOpen
  \bibfield  {author} {\bibinfo {author} {\bibfnamefont {Y.}~\bibnamefont {Wang}}, \bibinfo {author} {\bibfnamefont {Q.}~\bibnamefont {Zhang}}, \bibinfo {author} {\bibfnamefont {H.~P.}\ \bibnamefont {Astier}}, \bibinfo {author} {\bibfnamefont {C.}~\bibnamefont {Nickle}}, \bibinfo {author} {\bibfnamefont {S.}~\bibnamefont {Soni}}, \bibinfo {author} {\bibfnamefont {F.~A.}\ \bibnamefont {Alami}}, \bibinfo {author} {\bibfnamefont {A.}~\bibnamefont {Borrini}}, \bibinfo {author} {\bibfnamefont {Z.}~\bibnamefont {Zhang}}, \bibinfo {author} {\bibfnamefont {C.}~\bibnamefont {Honnigfort}}, \bibinfo {author} {\bibfnamefont {B.}~\bibnamefont {Braunschweig}}, \bibinfo {author} {\bibfnamefont {A.}~\bibnamefont {Leoncini}}, \bibinfo {author} {\bibfnamefont {D.~C.}\ \bibnamefont {Qi}}, \bibinfo {author} {\bibfnamefont {Y.}~\bibnamefont {Han}}, \bibinfo {author} {\bibfnamefont {E.}~\bibnamefont {del Barco}}, \bibinfo {author} {\bibfnamefont {D.}~\bibnamefont {Thompson}}, \ and\ \bibinfo {author} {\bibfnamefont {C.~A.}\
  \bibnamefont {Nijhuis}},\ }\bibfield  {title} {\enquote {\bibinfo {title} {{Dynamic molecular switches with hysteretic negative differential conductance emulating synaptic behaviour}},}\ }\href {\doibase 10.1038/s41563-022-01402-2} {\bibfield  {journal} {\bibinfo  {journal} {Nature Materials 2022 21:12}\ }\textbf {\bibinfo {volume} {21}},\ \bibinfo {pages} {1403--1411} (\bibinfo {year} {2022})}\BibitemShut {NoStop}%
\bibitem [{\citenamefont {Van De~Burgt}\ \emph {et~al.}(2018)\citenamefont {Van De~Burgt}, \citenamefont {Melianas}, \citenamefont {Keene}, \citenamefont {Malliaras},\ and\ \citenamefont {Salleo}}]{VanDeBurgt2018OrganicComputing}%
  \BibitemOpen
  \bibfield  {author} {\bibinfo {author} {\bibfnamefont {Y.}~\bibnamefont {Van De~Burgt}}, \bibinfo {author} {\bibfnamefont {A.}~\bibnamefont {Melianas}}, \bibinfo {author} {\bibfnamefont {S.~T.}\ \bibnamefont {Keene}}, \bibinfo {author} {\bibfnamefont {G.}~\bibnamefont {Malliaras}}, \ and\ \bibinfo {author} {\bibfnamefont {A.}~\bibnamefont {Salleo}},\ }\bibfield  {title} {\enquote {\bibinfo {title} {{Organic electronics for neuromorphic computing}},}\ }\href {\doibase 10.1038/s41928-018-0103-3} {\bibfield  {journal} {\bibinfo  {journal} {Nature Electronics}\ }\textbf {\bibinfo {volume} {1}},\ \bibinfo {pages} {386--397} (\bibinfo {year} {2018})}\BibitemShut {NoStop}%
\bibitem [{\citenamefont {Harikesh}\ \emph {et~al.}(2022)\citenamefont {Harikesh}, \citenamefont {Yang}, \citenamefont {Tu}, \citenamefont {Gerasimov}, \citenamefont {Dar}, \citenamefont {Armada-Moreira}, \citenamefont {Massetti}, \citenamefont {Kroon}, \citenamefont {Bliman}, \citenamefont {Olsson}, \citenamefont {Stavrinidou}, \citenamefont {Berggren},\ and\ \citenamefont {Fabiano}}]{Harikesh2022OrganicSpiking}%
  \BibitemOpen
  \bibfield  {author} {\bibinfo {author} {\bibfnamefont {P.~C.}\ \bibnamefont {Harikesh}}, \bibinfo {author} {\bibfnamefont {C.~Y.}\ \bibnamefont {Yang}}, \bibinfo {author} {\bibfnamefont {D.}~\bibnamefont {Tu}}, \bibinfo {author} {\bibfnamefont {J.~Y.}\ \bibnamefont {Gerasimov}}, \bibinfo {author} {\bibfnamefont {A.~M.}\ \bibnamefont {Dar}}, \bibinfo {author} {\bibfnamefont {A.}~\bibnamefont {Armada-Moreira}}, \bibinfo {author} {\bibfnamefont {M.}~\bibnamefont {Massetti}}, \bibinfo {author} {\bibfnamefont {R.}~\bibnamefont {Kroon}}, \bibinfo {author} {\bibfnamefont {D.}~\bibnamefont {Bliman}}, \bibinfo {author} {\bibfnamefont {R.}~\bibnamefont {Olsson}}, \bibinfo {author} {\bibfnamefont {E.}~\bibnamefont {Stavrinidou}}, \bibinfo {author} {\bibfnamefont {M.}~\bibnamefont {Berggren}}, \ and\ \bibinfo {author} {\bibfnamefont {S.}~\bibnamefont {Fabiano}},\ }\bibfield  {title} {\enquote {\bibinfo {title} {{Organic electrochemical neurons and synapses with ion mediated spiking}},}\ }\href {\doibase
  10.1038/s41467-022-28483-6} {\bibfield  {journal} {\bibinfo  {journal} {Nature Communications 2022 13:1}\ }\textbf {\bibinfo {volume} {13}},\ \bibinfo {pages} {1--9} (\bibinfo {year} {2022})}\BibitemShut {NoStop}%
\bibitem [{\citenamefont {Harikesh}\ \emph {et~al.}(2023)\citenamefont {Harikesh}, \citenamefont {Yang}, \citenamefont {Wu}, \citenamefont {Zhang}, \citenamefont {Donahue}, \citenamefont {Caravaca}, \citenamefont {Huang}, \citenamefont {Olofsson}, \citenamefont {Berggren}, \citenamefont {Tu},\ and\ \citenamefont {Fabiano}}]{Harikesh2023IonTunable}%
  \BibitemOpen
  \bibfield  {author} {\bibinfo {author} {\bibfnamefont {P.~C.}\ \bibnamefont {Harikesh}}, \bibinfo {author} {\bibfnamefont {C.~Y.}\ \bibnamefont {Yang}}, \bibinfo {author} {\bibfnamefont {H.~Y.}\ \bibnamefont {Wu}}, \bibinfo {author} {\bibfnamefont {S.}~\bibnamefont {Zhang}}, \bibinfo {author} {\bibfnamefont {M.~J.}\ \bibnamefont {Donahue}}, \bibinfo {author} {\bibfnamefont {A.~S.}\ \bibnamefont {Caravaca}}, \bibinfo {author} {\bibfnamefont {J.~D.}\ \bibnamefont {Huang}}, \bibinfo {author} {\bibfnamefont {P.~S.}\ \bibnamefont {Olofsson}}, \bibinfo {author} {\bibfnamefont {M.}~\bibnamefont {Berggren}}, \bibinfo {author} {\bibfnamefont {D.}~\bibnamefont {Tu}}, \ and\ \bibinfo {author} {\bibfnamefont {S.}~\bibnamefont {Fabiano}},\ }\bibfield  {title} {\enquote {\bibinfo {title} {{Ion-tunable antiambipolarity in mixed ion-electron conducting polymers enables biorealistic organic electrochemical neurons}},}\ }\href {\doibase 10.1038/s41563-022-01450-8} {\bibfield  {journal} {\bibinfo  {journal} {Nature
  Materials}\ }\textbf {\bibinfo {volume} {22}},\ \bibinfo {pages} {242--248} (\bibinfo {year} {2023})}\BibitemShut {NoStop}%
\bibitem [{\citenamefont {Luo}\ \emph {et~al.}(2023)\citenamefont {Luo}, \citenamefont {Shao}, \citenamefont {Ji}, \citenamefont {Chen}, \citenamefont {Wang}, \citenamefont {Wu}, \citenamefont {Kong}, \citenamefont {Guo}, \citenamefont {Wei}, \citenamefont {Zhao}, \citenamefont {Liu},\ and\ \citenamefont {Wei}}]{Luo2023HighlyDynamics}%
  \BibitemOpen
  \bibfield  {author} {\bibinfo {author} {\bibfnamefont {S.}~\bibnamefont {Luo}}, \bibinfo {author} {\bibfnamefont {L.}~\bibnamefont {Shao}}, \bibinfo {author} {\bibfnamefont {D.}~\bibnamefont {Ji}}, \bibinfo {author} {\bibfnamefont {Y.}~\bibnamefont {Chen}}, \bibinfo {author} {\bibfnamefont {X.}~\bibnamefont {Wang}}, \bibinfo {author} {\bibfnamefont {Y.}~\bibnamefont {Wu}}, \bibinfo {author} {\bibfnamefont {D.}~\bibnamefont {Kong}}, \bibinfo {author} {\bibfnamefont {M.}~\bibnamefont {Guo}}, \bibinfo {author} {\bibfnamefont {D.}~\bibnamefont {Wei}}, \bibinfo {author} {\bibfnamefont {Y.}~\bibnamefont {Zhao}}, \bibinfo {author} {\bibfnamefont {Y.}~\bibnamefont {Liu}}, \ and\ \bibinfo {author} {\bibfnamefont {D.}~\bibnamefont {Wei}},\ }\bibfield  {title} {\enquote {\bibinfo {title} {{Highly Bionic Neurotransmitter-Communicated Neurons Following Integrate-and-Fire Dynamics}},}\ }\href {\doibase 10.1021/ACS.NANOLETT.3C00799/ASSET/IMAGES/LARGE/NL3C00799{\_}0005.JPEG} {\bibfield  {journal} {\bibinfo  {journal} {Nano
  Letters}\ }\textbf {\bibinfo {volume} {23}},\ \bibinfo {pages} {4974--4982} (\bibinfo {year} {2023})}\BibitemShut {NoStop}%
\bibitem [{\citenamefont {Noy}\ and\ \citenamefont {Darling}(2023)}]{Noy2023NanofluidicSplash}%
  \BibitemOpen
  \bibfield  {author} {\bibinfo {author} {\bibfnamefont {A.}~\bibnamefont {Noy}}\ and\ \bibinfo {author} {\bibfnamefont {S.~B.}\ \bibnamefont {Darling}},\ }\bibfield  {title} {\enquote {\bibinfo {title} {{Nanofluidic computing makes a splash}},}\ }\href {\doibase 10.1126/science.adf6400} {\bibfield  {journal} {\bibinfo  {journal} {Science}\ }\textbf {\bibinfo {volume} {379}},\ \bibinfo {pages} {143--144} (\bibinfo {year} {2023})}\BibitemShut {NoStop}%
\bibitem [{\citenamefont {Noy}\ \emph {et~al.}(2023)\citenamefont {Noy}, \citenamefont {Li},\ and\ \citenamefont {Darling}}]{Noy2023FluidDevices}%
  \BibitemOpen
  \bibfield  {author} {\bibinfo {author} {\bibfnamefont {A.}~\bibnamefont {Noy}}, \bibinfo {author} {\bibfnamefont {Z.}~\bibnamefont {Li}}, \ and\ \bibinfo {author} {\bibfnamefont {S.~B.}\ \bibnamefont {Darling}},\ }\bibfield  {title} {\enquote {\bibinfo {title} {{Fluid learning: Mimicking brain computing with neuromorphic nanofluidic devices}},}\ }\href {\doibase 10.1016/J.NANTOD.2023.102043} {\bibfield  {journal} {\bibinfo  {journal} {Nano Today}\ }\textbf {\bibinfo {volume} {53}},\ \bibinfo {pages} {102043} (\bibinfo {year} {2023})}\BibitemShut {NoStop}%
\bibitem [{\citenamefont {Robin}\ \emph {et~al.}(2021)\citenamefont {Robin}, \citenamefont {Kavokine},\ and\ \citenamefont {Bocquet}}]{Robin2021ModelingSlits}%
  \BibitemOpen
  \bibfield  {author} {\bibinfo {author} {\bibfnamefont {P.}~\bibnamefont {Robin}}, \bibinfo {author} {\bibfnamefont {N.}~\bibnamefont {Kavokine}}, \ and\ \bibinfo {author} {\bibfnamefont {L.}~\bibnamefont {Bocquet}},\ }\bibfield  {title} {\enquote {\bibinfo {title} {{Modeling of emergent memory and voltage spiking in ionic transport through angstrom-scale slits}},}\ }\href {https://www.science.org} {\bibfield  {journal} {\bibinfo  {journal} {Science}\ }\textbf {\bibinfo {volume} {373}},\ \bibinfo {pages} {687--691} (\bibinfo {year} {2021})}\BibitemShut {NoStop}%
\bibitem [{\citenamefont {Robin}\ \emph {et~al.}(2023)\citenamefont {Robin}, \citenamefont {Emmerich}, \citenamefont {Ismail}, \citenamefont {Nigu{\`{e}}s}, \citenamefont {You}, \citenamefont {Nam}, \citenamefont {Keerthi}, \citenamefont {Siria}, \citenamefont {Geim}, \citenamefont {Radha},\ and\ \citenamefont {Bocquet}}]{Robin2023Long-termChannels}%
  \BibitemOpen
  \bibfield  {author} {\bibinfo {author} {\bibfnamefont {P.}~\bibnamefont {Robin}}, \bibinfo {author} {\bibfnamefont {T.}~\bibnamefont {Emmerich}}, \bibinfo {author} {\bibfnamefont {A.}~\bibnamefont {Ismail}}, \bibinfo {author} {\bibfnamefont {A.}~\bibnamefont {Nigu{\`{e}}s}}, \bibinfo {author} {\bibfnamefont {Y.}~\bibnamefont {You}}, \bibinfo {author} {\bibfnamefont {G.-h.}\ \bibnamefont {Nam}}, \bibinfo {author} {\bibfnamefont {A.}~\bibnamefont {Keerthi}}, \bibinfo {author} {\bibfnamefont {A.}~\bibnamefont {Siria}}, \bibinfo {author} {\bibfnamefont {A.~K.}\ \bibnamefont {Geim}}, \bibinfo {author} {\bibfnamefont {B.}~\bibnamefont {Radha}}, \ and\ \bibinfo {author} {\bibfnamefont {L.}~\bibnamefont {Bocquet}},\ }\bibfield  {title} {\enquote {\bibinfo {title} {{Long-term memory and synapse-like dynamics in two-dimensional nanofluidic channels}},}\ }\href {\doibase 10.1126/science.adc9931} {\bibfield  {journal} {\bibinfo  {journal} {Science}\ }\textbf {\bibinfo {volume} {379}},\ \bibinfo {pages} {161--167}
  (\bibinfo {year} {2023})}\BibitemShut {NoStop}%
\bibitem [{\citenamefont {Xiong}\ \emph {et~al.}(2023{\natexlab{a}})\citenamefont {Xiong}, \citenamefont {Li}, \citenamefont {He}, \citenamefont {Xie}, \citenamefont {Zong}, \citenamefont {Jiang}, \citenamefont {Ma}, \citenamefont {Wu}, \citenamefont {Fei}, \citenamefont {Yu},\ and\ \citenamefont {Mao}}]{Xiong2023NeuromorphicMemristor}%
  \BibitemOpen
  \bibfield  {author} {\bibinfo {author} {\bibfnamefont {T.}~\bibnamefont {Xiong}}, \bibinfo {author} {\bibfnamefont {C.}~\bibnamefont {Li}}, \bibinfo {author} {\bibfnamefont {X.}~\bibnamefont {He}}, \bibinfo {author} {\bibfnamefont {B.}~\bibnamefont {Xie}}, \bibinfo {author} {\bibfnamefont {J.}~\bibnamefont {Zong}}, \bibinfo {author} {\bibfnamefont {Y.}~\bibnamefont {Jiang}}, \bibinfo {author} {\bibfnamefont {W.}~\bibnamefont {Ma}}, \bibinfo {author} {\bibfnamefont {F.}~\bibnamefont {Wu}}, \bibinfo {author} {\bibfnamefont {J.}~\bibnamefont {Fei}}, \bibinfo {author} {\bibfnamefont {P.}~\bibnamefont {Yu}}, \ and\ \bibinfo {author} {\bibfnamefont {L.}~\bibnamefont {Mao}},\ }\bibfield  {title} {\enquote {\bibinfo {title} {{Neuromorphic functions with a polyelectrolyte-confined fluidic memristor}},}\ }\href {\doibase 10.1126/science.adc9150} {\bibfield  {journal} {\bibinfo  {journal} {Science}\ }\textbf {\bibinfo {volume} {379}},\ \bibinfo {pages} {156--161} (\bibinfo {year} {2023}{\natexlab{a}})}\BibitemShut
  {NoStop}%
\bibitem [{\citenamefont {Emmerich}\ \emph {et~al.}(2023)\citenamefont {Emmerich}, \citenamefont {Teng}, \citenamefont {Ronceray}, \citenamefont {Lopriore}, \citenamefont {Chiesa}, \citenamefont {Chernev}, \citenamefont {Artemov}, \citenamefont {Di~Ventra}, \citenamefont {Kis},\ and\ \citenamefont {Radenovic}}]{Emmerich2023IonicSwitches}%
  \BibitemOpen
  \bibfield  {author} {\bibinfo {author} {\bibfnamefont {T.}~\bibnamefont {Emmerich}}, \bibinfo {author} {\bibfnamefont {Y.}~\bibnamefont {Teng}}, \bibinfo {author} {\bibfnamefont {N.}~\bibnamefont {Ronceray}}, \bibinfo {author} {\bibfnamefont {E.}~\bibnamefont {Lopriore}}, \bibinfo {author} {\bibfnamefont {R.}~\bibnamefont {Chiesa}}, \bibinfo {author} {\bibfnamefont {A.}~\bibnamefont {Chernev}}, \bibinfo {author} {\bibfnamefont {V.}~\bibnamefont {Artemov}}, \bibinfo {author} {\bibfnamefont {M.}~\bibnamefont {Di~Ventra}}, \bibinfo {author} {\bibfnamefont {A.}~\bibnamefont {Kis}}, \ and\ \bibinfo {author} {\bibfnamefont {A.}~\bibnamefont {Radenovic}},\ }\bibfield  {title} {\enquote {\bibinfo {title} {{Ionic logic with highly asymmetric nanofluidic memristive switches}},}\ }\href {http://arxiv.org/abs/2306.07617} {\bibfield  {journal} {\bibinfo  {journal} {arXiv preprint}\ } (\bibinfo {year} {2023})}\BibitemShut {NoStop}%
\bibitem [{\citenamefont {Han}\ \emph {et~al.}(2023)\citenamefont {Han}, \citenamefont {Kim}, \citenamefont {Oh},\ and\ \citenamefont {Chung}}]{Han2023IontronicDissolution}%
  \BibitemOpen
  \bibfield  {author} {\bibinfo {author} {\bibfnamefont {S.~H.}\ \bibnamefont {Han}}, \bibinfo {author} {\bibfnamefont {S.~I.}\ \bibnamefont {Kim}}, \bibinfo {author} {\bibfnamefont {M.~A.}\ \bibnamefont {Oh}}, \ and\ \bibinfo {author} {\bibfnamefont {T.~D.}\ \bibnamefont {Chung}},\ }\bibfield  {title} {\enquote {\bibinfo {title} {{Iontronic analog of synaptic plasticity: Hydrogel-based ionic diode with chemical precipitation and dissolution}},}\ }\href {\doibase 10.1073/pnas.2211442120} {\bibfield  {journal} {\bibinfo  {journal} {Proceedings of the National Academy of Sciences of the United States of America}\ }\textbf {\bibinfo {volume} {120}},\ \bibinfo {pages} {e2211442120} (\bibinfo {year} {2023})}\BibitemShut {NoStop}%
\bibitem [{\citenamefont {Xie}\ \emph {et~al.}(2022)\citenamefont {Xie}, \citenamefont {Xiong}, \citenamefont {Li}, \citenamefont {Gao}, \citenamefont {Zong}, \citenamefont {Liu},\ and\ \citenamefont {Yu}}]{Xie2022PerspectiveApplication}%
  \BibitemOpen
  \bibfield  {author} {\bibinfo {author} {\bibfnamefont {B.}~\bibnamefont {Xie}}, \bibinfo {author} {\bibfnamefont {T.}~\bibnamefont {Xiong}}, \bibinfo {author} {\bibfnamefont {W.}~\bibnamefont {Li}}, \bibinfo {author} {\bibfnamefont {T.}~\bibnamefont {Gao}}, \bibinfo {author} {\bibfnamefont {J.}~\bibnamefont {Zong}}, \bibinfo {author} {\bibfnamefont {Y.}~\bibnamefont {Liu}}, \ and\ \bibinfo {author} {\bibfnamefont {P.}~\bibnamefont {Yu}},\ }\bibfield  {title} {\enquote {\bibinfo {title} {{Perspective on Nanofluidic Memristors: From Mechanism to Application}},}\ }\href {\doibase 10.1002/ASIA.202200682} {\bibfield  {journal} {\bibinfo  {journal} {Chemistry - An Asian Journal}\ }\textbf {\bibinfo {volume} {17}},\ \bibinfo {pages} {e202200682} (\bibinfo {year} {2022})}\BibitemShut {NoStop}%
\bibitem [{\citenamefont {Han}\ \emph {et~al.}(2022)\citenamefont {Han}, \citenamefont {Oh},\ and\ \citenamefont {Chung}}]{Han2022Iontronics:Applications}%
  \BibitemOpen
  \bibfield  {author} {\bibinfo {author} {\bibfnamefont {S.~H.}\ \bibnamefont {Han}}, \bibinfo {author} {\bibfnamefont {M.-A.}\ \bibnamefont {Oh}}, \ and\ \bibinfo {author} {\bibfnamefont {T.~D.}\ \bibnamefont {Chung}},\ }\bibfield  {title} {\enquote {\bibinfo {title} {{Iontronics: Aqueous ion-based engineering for bioinspired functionalities and applications}},}\ }\href {\doibase 10.1063/5.0089822} {\bibfield  {journal} {\bibinfo  {journal} {Chemical Physics Reviews}\ }\textbf {\bibinfo {volume} {3}},\ \bibinfo {pages} {031302} (\bibinfo {year} {2022})}\BibitemShut {NoStop}%
\bibitem [{\citenamefont {Bocquet}(2023)}]{Bocquet2023ConcludingMemories}%
  \BibitemOpen
  \bibfield  {author} {\bibinfo {author} {\bibfnamefont {L.}~\bibnamefont {Bocquet}},\ }\bibfield  {title} {\enquote {\bibinfo {title} {{Concluding remarks: Iontronics, from fundamentals to ion-controlled devices-Random access memories}},}\ }\href {\doibase 10.1039/d3fd00138e} {\  (\bibinfo {year} {2023}),\ 10.1039/d3fd00138e}\BibitemShut {NoStop}%
\bibitem [{\citenamefont {Xiong}\ \emph {et~al.}(2023{\natexlab{b}})\citenamefont {Xiong}, \citenamefont {Li}, \citenamefont {Yu},\ and\ \citenamefont {Mao}}]{Xiong2023FluidicDevices}%
  \BibitemOpen
  \bibfield  {author} {\bibinfo {author} {\bibfnamefont {T.}~\bibnamefont {Xiong}}, \bibinfo {author} {\bibfnamefont {W.}~\bibnamefont {Li}}, \bibinfo {author} {\bibfnamefont {P.}~\bibnamefont {Yu}}, \ and\ \bibinfo {author} {\bibfnamefont {L.}~\bibnamefont {Mao}},\ }\bibfield  {title} {\enquote {\bibinfo {title} {{Fluidic memristor: Bringing chemistry to neuromorphic devices}},}\ }\href {\doibase 10.1016/J.XINN.2023.100435} {\bibfield  {journal} {\bibinfo  {journal} {The Innovation}\ }\textbf {\bibinfo {volume} {4}},\ \bibinfo {pages} {100435} (\bibinfo {year} {2023}{\natexlab{b}})}\BibitemShut {NoStop}%
\bibitem [{\citenamefont {Kamsma}\ \emph {et~al.}(2023{\natexlab{b}})\citenamefont {Kamsma}, \citenamefont {Kim}, \citenamefont {Kim}, \citenamefont {Boon}, \citenamefont {Spitoni}, \citenamefont {Park},\ and\ \citenamefont {van Roij}}]{Kamsma2023Brain-inspiredNanochannels}%
  \BibitemOpen
  \bibfield  {author} {\bibinfo {author} {\bibfnamefont {T.~M.}\ \bibnamefont {Kamsma}}, \bibinfo {author} {\bibfnamefont {J.}~\bibnamefont {Kim}}, \bibinfo {author} {\bibfnamefont {K.}~\bibnamefont {Kim}}, \bibinfo {author} {\bibfnamefont {W.~Q.}\ \bibnamefont {Boon}}, \bibinfo {author} {\bibfnamefont {C.}~\bibnamefont {Spitoni}}, \bibinfo {author} {\bibfnamefont {J.}~\bibnamefont {Park}}, \ and\ \bibinfo {author} {\bibfnamefont {R.}~\bibnamefont {van Roij}},\ }\bibfield  {title} {\enquote {\bibinfo {title} {{Brain-inspired computing with fluidic iontronic nanochannels}},}\ }\href {https://arxiv.org/abs/2309.11438v1} {\bibfield  {journal} {\bibinfo  {journal} {arXiv}\ } (\bibinfo {year} {2023}{\natexlab{b}})}\BibitemShut {NoStop}%
\bibitem [{\citenamefont {Kamsma}\ \emph {et~al.}(2023{\natexlab{c}})\citenamefont {Kamsma}, \citenamefont {Boon}, \citenamefont {Spitoni},\ and\ \citenamefont {van Roij}}]{Kamsma2023UnveilingIontronics}%
  \BibitemOpen
  \bibfield  {author} {\bibinfo {author} {\bibfnamefont {T.~M.}\ \bibnamefont {Kamsma}}, \bibinfo {author} {\bibfnamefont {W.~Q.}\ \bibnamefont {Boon}}, \bibinfo {author} {\bibfnamefont {C.}~\bibnamefont {Spitoni}}, \ and\ \bibinfo {author} {\bibfnamefont {R.}~\bibnamefont {van Roij}},\ }\bibfield  {title} {\enquote {\bibinfo {title} {{Unveiling the capabilities of bipolar conical channels in neuromorphic iontronics}},}\ }\href {\doibase 10.1039/D3FD00022B} {\bibfield  {journal} {\bibinfo  {journal} {Faraday Discussions}\ }\textbf {\bibinfo {volume} {246}},\ \bibinfo {pages} {125--140} (\bibinfo {year} {2023}{\natexlab{c}})}\BibitemShut {NoStop}%
\bibitem [{\citenamefont {Wang}\ \emph {et~al.}(2012)\citenamefont {Wang}, \citenamefont {Kvetny}, \citenamefont {Liu}, \citenamefont {Brown}, \citenamefont {Li},\ and\ \citenamefont {Wang}}]{Wang2012TransmembraneTransport}%
  \BibitemOpen
  \bibfield  {author} {\bibinfo {author} {\bibfnamefont {D.}~\bibnamefont {Wang}}, \bibinfo {author} {\bibfnamefont {M.}~\bibnamefont {Kvetny}}, \bibinfo {author} {\bibfnamefont {J.}~\bibnamefont {Liu}}, \bibinfo {author} {\bibfnamefont {W.}~\bibnamefont {Brown}}, \bibinfo {author} {\bibfnamefont {Y.}~\bibnamefont {Li}}, \ and\ \bibinfo {author} {\bibfnamefont {G.}~\bibnamefont {Wang}},\ }\bibfield  {title} {\enquote {\bibinfo {title} {{Transmembrane potential across single conical nanopores and resulting memristive and memcapacitive ion transport}},}\ }\href {\doibase 10.1021/ja211142e} {\bibfield  {journal} {\bibinfo  {journal} {Journal of the American Chemical Society}\ }\textbf {\bibinfo {volume} {134}},\ \bibinfo {pages} {3651--3654} (\bibinfo {year} {2012})}\BibitemShut {NoStop}%
\bibitem [{\citenamefont {Ramirez}\ \emph {et~al.}(2023)\citenamefont {Ramirez}, \citenamefont {G{\'{o}}mez}, \citenamefont {Cervera}, \citenamefont {Mafe},\ and\ \citenamefont {Bisquert}}]{Ramirez2023SynapticalMemristors}%
  \BibitemOpen
  \bibfield  {author} {\bibinfo {author} {\bibfnamefont {P.}~\bibnamefont {Ramirez}}, \bibinfo {author} {\bibfnamefont {V.}~\bibnamefont {G{\'{o}}mez}}, \bibinfo {author} {\bibfnamefont {J.}~\bibnamefont {Cervera}}, \bibinfo {author} {\bibfnamefont {S.}~\bibnamefont {Mafe}}, \ and\ \bibinfo {author} {\bibfnamefont {J.}~\bibnamefont {Bisquert}},\ }\bibfield  {title} {\enquote {\bibinfo {title} {{Synaptical Tunability of Multipore Nanofluidic Memristors}},}\ }\href {\doibase 10.1021/ACS.JPCLETT.3C02796} {\bibfield  {journal} {\bibinfo  {journal} {The Journal of Physical Chemistry Letters}\ }\textbf {\bibinfo {volume} {14}},\ \bibinfo {pages} {10930--10934} (\bibinfo {year} {2023})}\BibitemShut {NoStop}%
\bibitem [{\citenamefont {Daiguji}\ \emph {et~al.}(2005)\citenamefont {Daiguji}, \citenamefont {Oka},\ and\ \citenamefont {Shirono}}]{Daiguji2005NanofluidicTransistor}%
  \BibitemOpen
  \bibfield  {author} {\bibinfo {author} {\bibfnamefont {H.}~\bibnamefont {Daiguji}}, \bibinfo {author} {\bibfnamefont {Y.}~\bibnamefont {Oka}}, \ and\ \bibinfo {author} {\bibfnamefont {K.}~\bibnamefont {Shirono}},\ }\bibfield  {title} {\enquote {\bibinfo {title} {{Nanofluidic diode and bipolar transistor}},}\ }\href {\doibase 10.1021/NL051646Y} {\bibfield  {journal} {\bibinfo  {journal} {Nano Letters}\ }\textbf {\bibinfo {volume} {5}},\ \bibinfo {pages} {2274--2280} (\bibinfo {year} {2005})}\BibitemShut {NoStop}%
\bibitem [{\citenamefont {Vlassiouk}\ and\ \citenamefont {Siwy}(2007)}]{Vlassiouk2007NanofluidicDiode}%
  \BibitemOpen
  \bibfield  {author} {\bibinfo {author} {\bibfnamefont {I.}~\bibnamefont {Vlassiouk}}\ and\ \bibinfo {author} {\bibfnamefont {Z.~S.}\ \bibnamefont {Siwy}},\ }\bibfield  {title} {\enquote {\bibinfo {title} {{Nanofluidic diode}},}\ }\href {\doibase 10.1021/NL062924B} {\bibfield  {journal} {\bibinfo  {journal} {Nano Letters}\ }\textbf {\bibinfo {volume} {7}},\ \bibinfo {pages} {552--556} (\bibinfo {year} {2007})}\BibitemShut {NoStop}%
\bibitem [{\citenamefont {Strathmann}\ \emph {et~al.}(1997)\citenamefont {Strathmann}, \citenamefont {Krol}, \citenamefont {Rapp},\ and\ \citenamefont {Eigenberger}}]{Strathmann1997LimitingMembranes}%
  \BibitemOpen
  \bibfield  {author} {\bibinfo {author} {\bibfnamefont {H.}~\bibnamefont {Strathmann}}, \bibinfo {author} {\bibfnamefont {J.~J.}\ \bibnamefont {Krol}}, \bibinfo {author} {\bibfnamefont {H.~J.}\ \bibnamefont {Rapp}}, \ and\ \bibinfo {author} {\bibfnamefont {G.}~\bibnamefont {Eigenberger}},\ }\bibfield  {title} {\enquote {\bibinfo {title} {{Limiting current density and water dissociation in bipolar membranes}},}\ }\href {\doibase 10.1016/S0376-7388(96)00185-8} {\bibfield  {journal} {\bibinfo  {journal} {Journal of Membrane Science}\ }\textbf {\bibinfo {volume} {125}},\ \bibinfo {pages} {123--142} (\bibinfo {year} {1997})}\BibitemShut {NoStop}%
\bibitem [{\citenamefont {Montes De~Oca}\ \emph {et~al.}(2022)\citenamefont {Montes De~Oca}, \citenamefont {Dhanasekaran}, \citenamefont {C{\'{o}}rdoba}, \citenamefont {Darling},\ and\ \citenamefont {De~Pablo}}]{MontesDeOca2022IonicSimulation}%
  \BibitemOpen
  \bibfield  {author} {\bibinfo {author} {\bibfnamefont {J.~M.}\ \bibnamefont {Montes De~Oca}}, \bibinfo {author} {\bibfnamefont {J.}~\bibnamefont {Dhanasekaran}}, \bibinfo {author} {\bibfnamefont {A.}~\bibnamefont {C{\'{o}}rdoba}}, \bibinfo {author} {\bibfnamefont {S.~B.}\ \bibnamefont {Darling}}, \ and\ \bibinfo {author} {\bibfnamefont {J.~J.}\ \bibnamefont {De~Pablo}},\ }\bibfield  {title} {\enquote {\bibinfo {title} {{Ionic Transport in Electrostatic Janus Membranes. An Explicit Solvent Molecular Dynamic Simulation}},}\ }\href {\doibase 10.1021/ACSNANO.1C07706/SUPPL{\_}FILE/NN1C07706{\_}SI{\_}001.PDF} {\bibfield  {journal} {\bibinfo  {journal} {ACS Nano}\ }\textbf {\bibinfo {volume} {16}},\ \bibinfo {pages} {3768--3775} (\bibinfo {year} {2022})}\BibitemShut {NoStop}%
\bibitem [{\citenamefont {C{\'{o}}rdoba}\ \emph {et~al.}(2023)\citenamefont {C{\'{o}}rdoba}, \citenamefont {Montes De~Oca}, \citenamefont {Dhanasekaran}, \citenamefont {Darling~Abd},\ and\ \citenamefont {De~Pablo}}]{Cordoba2023CurrentNanopores}%
  \BibitemOpen
  \bibfield  {author} {\bibinfo {author} {\bibfnamefont {A.}~\bibnamefont {C{\'{o}}rdoba}}, \bibinfo {author} {\bibfnamefont {J.~M.}\ \bibnamefont {Montes De~Oca}}, \bibinfo {author} {\bibfnamefont {J.}~\bibnamefont {Dhanasekaran}}, \bibinfo {author} {\bibfnamefont {S.~B.}\ \bibnamefont {Darling~Abd}}, \ and\ \bibinfo {author} {\bibfnamefont {J.~J.}\ \bibnamefont {De~Pablo}},\ }\bibfield  {title} {\enquote {\bibinfo {title} {{Current rectification by nanoparticles in bipolar nanopores}},}\ }\href {\doibase 10.1039/d2me00187j} {\bibfield  {journal} {\bibinfo  {journal} {Cite this: Mol. Syst. Des. Eng}\ }\textbf {\bibinfo {volume} {8}},\ \bibinfo {pages} {289} (\bibinfo {year} {2023})}\BibitemShut {NoStop}%
\bibitem [{\citenamefont {Huang}\ \emph {et~al.}(2018)\citenamefont {Huang}, \citenamefont {Kong}, \citenamefont {Wen}, \citenamefont {Jiang}, \citenamefont {Huang}, \citenamefont {Kong}, \citenamefont {Wen},\ and\ \citenamefont {Jiang}}]{Huang2018BioinspiredBipolar}%
  \BibitemOpen
  \bibfield  {author} {\bibinfo {author} {\bibfnamefont {X.}~\bibnamefont {Huang}}, \bibinfo {author} {\bibfnamefont {X.-Y.}\ \bibnamefont {Kong}}, \bibinfo {author} {\bibfnamefont {L.}~\bibnamefont {Wen}}, \bibinfo {author} {\bibfnamefont {L.}~\bibnamefont {Jiang}}, \bibinfo {author} {\bibfnamefont {X.}~\bibnamefont {Huang}}, \bibinfo {author} {\bibfnamefont {X.-y.}\ \bibnamefont {Kong}}, \bibinfo {author} {\bibfnamefont {L.}~\bibnamefont {Wen}}, \ and\ \bibinfo {author} {\bibfnamefont {L.}~\bibnamefont {Jiang}},\ }\bibfield  {title} {\enquote {\bibinfo {title} {{Bioinspired Ionic Diodes: From Unipolar to Bipolar}},}\ }\href {\doibase 10.1002/ADFM.201801079} {\bibfield  {journal} {\bibinfo  {journal} {Advanced Functional Materials}\ }\textbf {\bibinfo {volume} {28}},\ \bibinfo {pages} {1801079} (\bibinfo {year} {2018})}\BibitemShut {NoStop}%
\bibitem [{\citenamefont {Yang}\ \emph {et~al.}(2018)\citenamefont {Yang}, \citenamefont {Xie}, \citenamefont {Hou}, \citenamefont {Cheetham}, \citenamefont {Chen}, \citenamefont {Darling}, \citenamefont {Yang}, \citenamefont {Darling}, \citenamefont {Xie}, \citenamefont {Hou}, \citenamefont {Cheetham},\ and\ \citenamefont {Chen}}]{Yang2018JanusEfficiency}%
  \BibitemOpen
  \bibfield  {author} {\bibinfo {author} {\bibfnamefont {H.-C.}\ \bibnamefont {Yang}}, \bibinfo {author} {\bibfnamefont {Y.}~\bibnamefont {Xie}}, \bibinfo {author} {\bibfnamefont {J.}~\bibnamefont {Hou}}, \bibinfo {author} {\bibfnamefont {A.~K.}\ \bibnamefont {Cheetham}}, \bibinfo {author} {\bibfnamefont {V.}~\bibnamefont {Chen}}, \bibinfo {author} {\bibfnamefont {S.~B.}\ \bibnamefont {Darling}}, \bibinfo {author} {\bibfnamefont {H.-C.}\ \bibnamefont {Yang}}, \bibinfo {author} {\bibfnamefont {S.~B.}\ \bibnamefont {Darling}}, \bibinfo {author} {\bibfnamefont {Y.}~\bibnamefont {Xie}}, \bibinfo {author} {\bibfnamefont {J.}~\bibnamefont {Hou}}, \bibinfo {author} {\bibfnamefont {A.~K.}\ \bibnamefont {Cheetham}}, \ and\ \bibinfo {author} {\bibfnamefont {V.}~\bibnamefont {Chen}},\ }\bibfield  {title} {\enquote {\bibinfo {title} {{Janus Membranes: Creating Asymmetry for Energy Efficiency}},}\ }\href {\doibase 10.1002/ADMA.201801495} {\bibfield  {journal} {\bibinfo  {journal} {Advanced Materials}\ }\textbf {\bibinfo
  {volume} {30}},\ \bibinfo {pages} {1801495} (\bibinfo {year} {2018})}\BibitemShut {NoStop}%
\bibitem [{\citenamefont {Yan}\ \emph {et~al.}(2021)\citenamefont {Yan}, \citenamefont {Yang}, \citenamefont {Zhang}, \citenamefont {Wu}, \citenamefont {Cheng}, \citenamefont {Darling},\ and\ \citenamefont {Shao}}]{Yan2021PorousFunctionality}%
  \BibitemOpen
  \bibfield  {author} {\bibinfo {author} {\bibfnamefont {L.}~\bibnamefont {Yan}}, \bibinfo {author} {\bibfnamefont {X.}~\bibnamefont {Yang}}, \bibinfo {author} {\bibfnamefont {Y.}~\bibnamefont {Zhang}}, \bibinfo {author} {\bibfnamefont {Y.}~\bibnamefont {Wu}}, \bibinfo {author} {\bibfnamefont {Z.}~\bibnamefont {Cheng}}, \bibinfo {author} {\bibfnamefont {S.~B.}\ \bibnamefont {Darling}}, \ and\ \bibinfo {author} {\bibfnamefont {L.}~\bibnamefont {Shao}},\ }\bibfield  {title} {\enquote {\bibinfo {title} {{Porous Janus materials with unique asymmetries and functionality}},}\ }\href {\doibase 10.1016/J.MATTOD.2021.07.001} {\bibfield  {journal} {\bibinfo  {journal} {Materials Today}\ }\textbf {\bibinfo {volume} {51}},\ \bibinfo {pages} {626--647} (\bibinfo {year} {2021})}\BibitemShut {NoStop}%
\bibitem [{\citenamefont {Gentet}\ \emph {et~al.}(2000)\citenamefont {Gentet}, \citenamefont {Stuart},\ and\ \citenamefont {Clements}}]{Gentet2000DirectNeurons}%
  \BibitemOpen
  \bibfield  {author} {\bibinfo {author} {\bibfnamefont {L.~J.}\ \bibnamefont {Gentet}}, \bibinfo {author} {\bibfnamefont {G.~J.}\ \bibnamefont {Stuart}}, \ and\ \bibinfo {author} {\bibfnamefont {J.~D.}\ \bibnamefont {Clements}},\ }\bibfield  {title} {\enquote {\bibinfo {title} {{Direct Measurement of Specific Membrane Capacitance in Neurons}},}\ }\href {\doibase 10.1016/S0006-3495(00)76293-X} {\bibfield  {journal} {\bibinfo  {journal} {Biophysical Journal}\ }\textbf {\bibinfo {volume} {79}},\ \bibinfo {pages} {314--320} (\bibinfo {year} {2000})}\BibitemShut {NoStop}%
\bibitem [{\citenamefont {Major}\ \emph {et~al.}(1994)\citenamefont {Major}, \citenamefont {Larkman}, \citenamefont {Jonas}, \citenamefont {Sakmann},\ and\ \citenamefont {Jack}}]{Major1994DetailedSlices}%
  \BibitemOpen
  \bibfield  {author} {\bibinfo {author} {\bibfnamefont {G.}~\bibnamefont {Major}}, \bibinfo {author} {\bibfnamefont {A.~U.}\ \bibnamefont {Larkman}}, \bibinfo {author} {\bibfnamefont {P.}~\bibnamefont {Jonas}}, \bibinfo {author} {\bibfnamefont {B.}~\bibnamefont {Sakmann}}, \ and\ \bibinfo {author} {\bibfnamefont {J.~J.~B.}\ \bibnamefont {Jack}},\ }\bibfield  {title} {\enquote {\bibinfo {title} {{Detailed passive cable models of whole-cell recorded CA3 pyramidal neurons in rat hippocampal slices}},}\ }\href {\doibase 10.1523/JNEUROSCI.14-08-04613.1994} {\bibfield  {journal} {\bibinfo  {journal} {Journal of Neuroscience}\ }\textbf {\bibinfo {volume} {14}},\ \bibinfo {pages} {4613--4638} (\bibinfo {year} {1994})}\BibitemShut {NoStop}%
\bibitem [{\citenamefont {{L. Squire, D. Berg, F. Bloom, S. du Lac, A. Ghosh, N. Spitzer}}(2008)}]{fundNeuroTrain}%
  \BibitemOpen
  \bibfield  {author} {\bibinfo {author} {\bibnamefont {{L. Squire, D. Berg, F. Bloom, S. du Lac, A. Ghosh, N. Spitzer}}},\ }\href@noop {} {\emph {\bibinfo {title} {Fundamental Neuroscience}}},\ \bibinfo {edition} {3rd}\ ed.\ (\bibinfo  {publisher} {Academic Press},\ \bibinfo {year} {2008})\ Chap.~\bibinfo {chapter} {6}\BibitemShut {NoStop}%
\bibitem [{\citenamefont {Kamsma}\ \emph {et~al.}(2024)\citenamefont {Kamsma}, \citenamefont {van Roij},\ and\ \citenamefont {Spitoni}}]{kamsma2023math}%
  \BibitemOpen
  \bibfield  {author} {\bibinfo {author} {\bibfnamefont {T.~M.}\ \bibnamefont {Kamsma}}, \bibinfo {author} {\bibfnamefont {R.}~\bibnamefont {van Roij}}, \ and\ \bibinfo {author} {\bibfnamefont {C.}~\bibnamefont {Spitoni}},\ }\bibfield  {title} {\enquote {\bibinfo {title} {A simple mathematical theory for simple volatile memristors and their spiking circuits},}\ }\href {\doibase 10.13140/RG.2.2.13242.40640} {\bibfield  {journal} {\bibinfo  {journal} {ResearchGate preprint}\ } (\bibinfo {year} {2024}),\ 10.13140/RG.2.2.13242.40640}\BibitemShut {NoStop}%
\bibitem [{\citenamefont {Chay}(1983)}]{Chay1983EyringOscillations}%
  \BibitemOpen
  \bibfield  {author} {\bibinfo {author} {\bibfnamefont {T.~R.}\ \bibnamefont {Chay}},\ }\bibfield  {title} {\enquote {\bibinfo {title} {{Eyring rate theory in excitable membranes: Application to neuronal oscillations}},}\ }\href {\doibase 10.1021/J100238A043} {\bibfield  {journal} {\bibinfo  {journal} {Journal of Physical Chemistry}\ }\textbf {\bibinfo {volume} {87}},\ \bibinfo {pages} {2935--2940} (\bibinfo {year} {1983})}\BibitemShut {NoStop}%
\bibitem [{\citenamefont {Chay}(1985)}]{Chay1985ChaosCell}%
  \BibitemOpen
  \bibfield  {author} {\bibinfo {author} {\bibfnamefont {T.~R.}\ \bibnamefont {Chay}},\ }\bibfield  {title} {\enquote {\bibinfo {title} {{Chaos in a three-variable model of an excitable cell}},}\ }\href@noop {} {\ ,\ \bibinfo {pages} {233--242} (\bibinfo {year} {1985})}\BibitemShut {NoStop}%
\bibitem [{\citenamefont {Xu}\ \emph {et~al.}(2020)\citenamefont {Xu}, \citenamefont {Tan}, \citenamefont {Zhu}, \citenamefont {Bao}, \citenamefont {Hu},\ and\ \citenamefont {Bao}}]{Xu2020BifurcationsCircuit}%
  \BibitemOpen
  \bibfield  {author} {\bibinfo {author} {\bibfnamefont {Q.}~\bibnamefont {Xu}}, \bibinfo {author} {\bibfnamefont {X.}~\bibnamefont {Tan}}, \bibinfo {author} {\bibfnamefont {D.}~\bibnamefont {Zhu}}, \bibinfo {author} {\bibfnamefont {H.}~\bibnamefont {Bao}}, \bibinfo {author} {\bibfnamefont {Y.}~\bibnamefont {Hu}}, \ and\ \bibinfo {author} {\bibfnamefont {B.}~\bibnamefont {Bao}},\ }\bibfield  {title} {\enquote {\bibinfo {title} {{Bifurcations to bursting and spiking in the Chay neuron and their validation in a digital circuit}},}\ }\href {\doibase 10.1016/J.CHAOS.2020.110353} {\bibfield  {journal} {\bibinfo  {journal} {Chaos, Solitons {\&} Fractals}\ }\textbf {\bibinfo {volume} {141}},\ \bibinfo {pages} {110353} (\bibinfo {year} {2020})}\BibitemShut {NoStop}%
\bibitem [{\citenamefont {Riza~Putra}\ \emph {et~al.}(2021)\citenamefont {Riza~Putra}, \citenamefont {Tshwenya}, \citenamefont {Buckingham}, \citenamefont {Chen}, \citenamefont {Jeremiah~Aoki}, \citenamefont {Mathwig}, \citenamefont {Arotiba}, \citenamefont {Thompson}, \citenamefont {Li},\ and\ \citenamefont {Marken}}]{RizaPutra2021MicroscaleOverview}%
  \BibitemOpen
  \bibfield  {author} {\bibinfo {author} {\bibfnamefont {B.}~\bibnamefont {Riza~Putra}}, \bibinfo {author} {\bibfnamefont {L.}~\bibnamefont {Tshwenya}}, \bibinfo {author} {\bibfnamefont {M.~A.}\ \bibnamefont {Buckingham}}, \bibinfo {author} {\bibfnamefont {J.}~\bibnamefont {Chen}}, \bibinfo {author} {\bibfnamefont {K.}~\bibnamefont {Jeremiah~Aoki}}, \bibinfo {author} {\bibfnamefont {K.}~\bibnamefont {Mathwig}}, \bibinfo {author} {\bibfnamefont {O.~A.}\ \bibnamefont {Arotiba}}, \bibinfo {author} {\bibfnamefont {A.~K.}\ \bibnamefont {Thompson}}, \bibinfo {author} {\bibfnamefont {Z.}~\bibnamefont {Li}}, \ and\ \bibinfo {author} {\bibfnamefont {F.}~\bibnamefont {Marken}},\ }\bibfield  {title} {\enquote {\bibinfo {title} {{Microscale Ionic Diodes: An Overview}},}\ }\href {\doibase 10.1002/ELAN.202060614} {\bibfield  {journal} {\bibinfo  {journal} {Electroanalysis}\ }\textbf {\bibinfo {volume} {33}},\ \bibinfo {pages} {1398--1418} (\bibinfo {year} {2021})}\BibitemShut {NoStop}%
\bibitem [{\citenamefont {Cheng}\ and\ \citenamefont {Guo}(2010)}]{Cheng2010NanofluidicDiodes}%
  \BibitemOpen
  \bibfield  {author} {\bibinfo {author} {\bibfnamefont {L.-J.}\ \bibnamefont {Cheng}}\ and\ \bibinfo {author} {\bibfnamefont {L.~J.}\ \bibnamefont {Guo}},\ }\bibfield  {title} {\enquote {\bibinfo {title} {{Nanofluidic diodes}},}\ }\href {\doibase 10.1039/B822554K} {\bibfield  {journal} {\bibinfo  {journal} {Chemical Society Reviews}\ }\textbf {\bibinfo {volume} {39}},\ \bibinfo {pages} {923--938} (\bibinfo {year} {2010})}\BibitemShut {NoStop}%
\bibitem [{\citenamefont {Lan}\ \emph {et~al.}(2016)\citenamefont {Lan}, \citenamefont {Edwards}, \citenamefont {Luo}, \citenamefont {Perera}, \citenamefont {Wu}, \citenamefont {Martin},\ and\ \citenamefont {White}}]{Lan2016Voltage-RectifiedNanopores}%
  \BibitemOpen
  \bibfield  {author} {\bibinfo {author} {\bibfnamefont {W.~J.}\ \bibnamefont {Lan}}, \bibinfo {author} {\bibfnamefont {M.~A.}\ \bibnamefont {Edwards}}, \bibinfo {author} {\bibfnamefont {L.}~\bibnamefont {Luo}}, \bibinfo {author} {\bibfnamefont {R.~T.}\ \bibnamefont {Perera}}, \bibinfo {author} {\bibfnamefont {X.}~\bibnamefont {Wu}}, \bibinfo {author} {\bibfnamefont {C.~R.}\ \bibnamefont {Martin}}, \ and\ \bibinfo {author} {\bibfnamefont {H.~S.}\ \bibnamefont {White}},\ }\bibfield  {title} {\enquote {\bibinfo {title} {{Voltage-Rectified Current and Fluid Flow in Conical Nanopores}},}\ }\href {\doibase 10.1021/ACS.ACCOUNTS.6B00395/ASSET/IMAGES/LARGE/AR-2016-00395C{\_}0011.JPEG} {\bibfield  {journal} {\bibinfo  {journal} {Accounts of Chemical Research}\ }\textbf {\bibinfo {volume} {49}},\ \bibinfo {pages} {2605--2613} (\bibinfo {year} {2016})}\BibitemShut {NoStop}%
\bibitem [{\citenamefont {Kim}\ \emph {et~al.}(2022)\citenamefont {Kim}, \citenamefont {Jeon}, \citenamefont {Wang}, \citenamefont {Chang},\ and\ \citenamefont {Park}}]{Kim2022AsymmetricDetection}%
  \BibitemOpen
  \bibfield  {author} {\bibinfo {author} {\bibfnamefont {J.}~\bibnamefont {Kim}}, \bibinfo {author} {\bibfnamefont {J.}~\bibnamefont {Jeon}}, \bibinfo {author} {\bibfnamefont {C.}~\bibnamefont {Wang}}, \bibinfo {author} {\bibfnamefont {G.~T.}\ \bibnamefont {Chang}}, \ and\ \bibinfo {author} {\bibfnamefont {J.}~\bibnamefont {Park}},\ }\bibfield  {title} {\enquote {\bibinfo {title} {{Asymmetric Nanochannel Network-Based Bipolar Ionic Diode for Enhanced Heavy Metal Ion Detection}},}\ }\href {\doibase 10.1021/acsnano.2c02016} {\bibfield  {journal} {\bibinfo  {journal} {ACS Nano}\ } (\bibinfo {year} {2022}),\ 10.1021/acsnano.2c02016}\BibitemShut {NoStop}%
\bibitem [{\citenamefont {Choi}\ \emph {et~al.}(2016)\citenamefont {Choi}, \citenamefont {Wang}, \citenamefont {Chang},\ and\ \citenamefont {Park}}]{Choi2016HighMembrane}%
  \BibitemOpen
  \bibfield  {author} {\bibinfo {author} {\bibfnamefont {E.}~\bibnamefont {Choi}}, \bibinfo {author} {\bibfnamefont {C.}~\bibnamefont {Wang}}, \bibinfo {author} {\bibfnamefont {G.~T.}\ \bibnamefont {Chang}}, \ and\ \bibinfo {author} {\bibfnamefont {J.}~\bibnamefont {Park}},\ }\bibfield  {title} {\enquote {\bibinfo {title} {{High Current Ionic Diode Using Homogeneously Charged Asymmetric Nanochannel Network Membrane}},}\ }\href {\doibase 10.1021/acs.nanolett.5b04246} {\bibfield  {journal} {\bibinfo  {journal} {Nano Letters}\ }\textbf {\bibinfo {volume} {16}},\ \bibinfo {pages} {2189--2197} (\bibinfo {year} {2016})}\BibitemShut {NoStop}%
\end{thebibliography}

%

\end{document}


\title{Supplemental Material for: Advanced iontronic spiking modes with multiscale diffusive dynamics in a fluidic circuit}

\author{T. M. Kamsma}
\affiliation{Institute for Theoretical Physics, Utrecht University,  Princetonplein 5, 3584 CC Utrecht, The Netherlands}
\affiliation{Mathematical Institute, Utrecht University, Budapestlaan 6, 3584 CD Utrecht, The Netherlands}
\author{E. A. Rossing}
\affiliation{Institute for Theoretical Physics, Utrecht University,  Princetonplein 5, 3584 CC Utrecht, The Netherlands}
\author{C. Spitoni}
\affiliation{Mathematical Institute, Utrecht University, Budapestlaan 6, 3584 CD Utrecht, The Netherlands}
\author{R. van Roij}
\affiliation{Institute for Theoretical Physics, Utrecht University,  Princetonplein 5, 3584 CC Utrecht, The Netherlands}
\maketitle
\setcounter{figure}{0}
\renewcommand{\figurename}{Figure}
\renewcommand{\thefigure}{S\arabic{figure}}

\setcounter{equation}{0}
\renewcommand{\theequation}{S\arabic{equation}}

\section{Bipolar channel steady-state conductance}
\begin{figure}[ht!]
		\centering
		\includegraphics[width=0.5\textwidth]{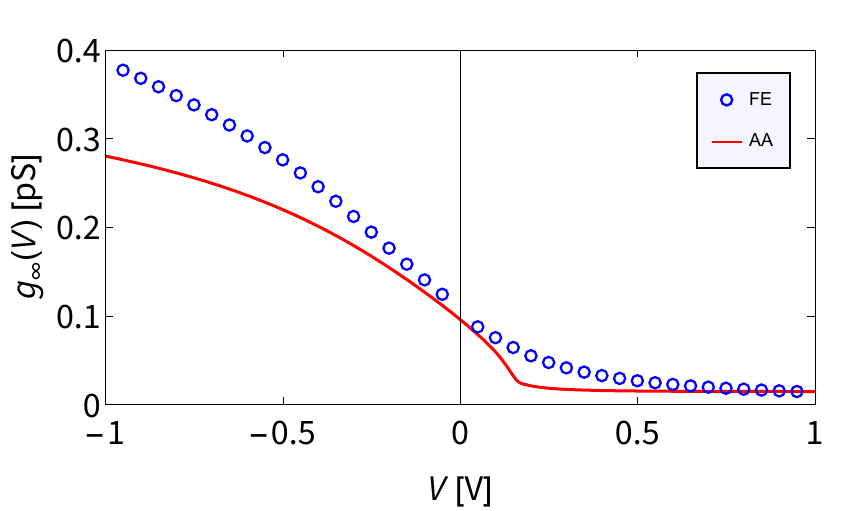}
		\caption{The voltage dependence of the steady-state conductance $g_{\infty}(V)$ as predicted by our analytic approximation of Eq.~(3) (red) and by finite-element calculations of the full Poisson-Nernst-Planck-Stokes equations (blue), as detailed in Ref.~\cite{Kamsma2023UnveilingIontronics}, for a bipolar channel of length $L=10\text{ }\mu$m, base radius $R_{\mathrm{b}}=200$ nm, tip radius $R_{\mathrm{t}}=50$ nm, and salt concentration $\rho_{\mathrm{b}}=2$ mM.}
		\label{fig:gplot}
\end{figure}

In Fig.~\ref{fig:gplot} we compare the steady-state conductance predicted by our analytical approximation (AA, red) of Eq.~(3) from the main text with finite-element (FE, blue) calculations of the full Poisson-Nernst-Planck-Stokes equations. The precise details of these FE calculations are detailed in Ref.~\cite{Kamsma2023UnveilingIontronics}. When compared to finite-element calculations we see that the analytical approximation of Eq.~(3), directly derived from the underlying Poisson-Nernst-Planck-Stokes equations, yields reasonable agreement, however it underestimates the increase in conductance for negative voltages while it overestimates the decrease in conductance for positive voltages. Eq.~(3) also predicts a relatively sharp conductance drop when concentrations start to approach the imposed minimum salt concentration of $0.2\rho_{\mathrm{b}}$ at around 0.15 V in Fig.\ref{fig:gplot},  which complicates the numerical solutions of the Kirchhoff equations of the main text. With a third-order interpolation of Eq.~(3) this sharp drop can be somewhat smoothed over by choosing intervals $\gtrsim 0.025$ V, in our case in the regime from -0.3125 to 0.3125 V. In the future, a more sophisticated theoretical model of individual channels should address the physics that underlies this detail, which involves a surface contribution to the channel conductance.

\section{Unipolar surface charge conical channel}
The tonic bursting described in the main text was also found to emerge from an iontronic circuit containing unipolar (UP) conical channels with a homogeneous surface charge, rather than bipolar (BP) surface charge channels as considered in the main text. These UP channels were found to be memristors that facilitated tonic and phasic spiking upon coupling them in an iontronic circuit containing three UP channels and a capacitor \cite{Kamsma2023IontronicMemristors}. Here we use essentially the same parameters and theoretical model used for tonic and phasic spiking in Ref.~\cite{Kamsma2023IontronicMemristors} as a basis to also find tonic bursting. That is, all UP channels have base and tip radii $R_{\mathrm{b}}=200\text{ nm}$ and $R_{\mathrm{t}}=R_{\mathrm{b}}-\Delta R=50\text{ nm}$, respectively, such that the channel radius is described by $R_i(x)=R_{\mathrm{b}}-x\Delta R/L_i$ for positions $x\in\left[0,L_i\right]$ in the channel. Here $L_i$ represents the channel length, which varies for different channels in the circuit. The channel connects two reservoirs containing an aqueous 1:1 electrolyte with ionic bulk concentration $\rho_{\mathrm{b}}=0.1$ mM (so substantially lower than considered in the main text)  and with the viscosity $\eta=1.01\text{ mPa}\cdot\text{s}$ and the electric permittivity $\epsilon=0.71\text{ nF}\cdot\text{m}^{-1}$ of water. We assume a uniform surface charge density $e\sigma=-0.0015\;e\text{nm}^{-2}$ on the channel walls, resulting in a surface potential $\psi_0\approx -10\text{ mV}$ and an electric double layer that screens the surface charge with Debye length $\lambda_{\mathrm{D}}\approx30\text{ nm}$. At the tip the Debye length is not much smaller than the channel radius, which does not fully satisfy the assumption of a small Debye length compared to the channel radius made in the theoretical framework we use to model the channel conductance \cite{Boon2023CoulombicKinetics}. However, it was shown that the dynamic conductance properties are still reasonably well described by the theoretical model we lay out below \cite{Kamsma2023IontronicMemristors}. The ions in this instance are assumed to all have diffusion coefficients $D=1.75\text{ }\mu\text{m}^2\text{ms}^{-1}$. The electro-osmotic volumetric fluid flow rate $Q_i(V_i)$ that is driven by an applied voltage $V_i$ over the channel is accurately represented by its linear-response approximation $Q_i(V_i)=-\pi R_{\mathrm{t}}R_{\mathrm{b}}\epsilon\psi_0V_i/(\eta L_i)$, and is conveniently characterised by the P\'{e}clet number at the narrow end $\text{Pe}_i(V_i)=Q_i(V_i)L_i/(D\pi R_{\mathrm{t}}^2)$ \cite{Kamsma2023IontronicMemristors}.

\begin{figure*}[ht]
		\centering
		\includegraphics[width=1\textwidth]{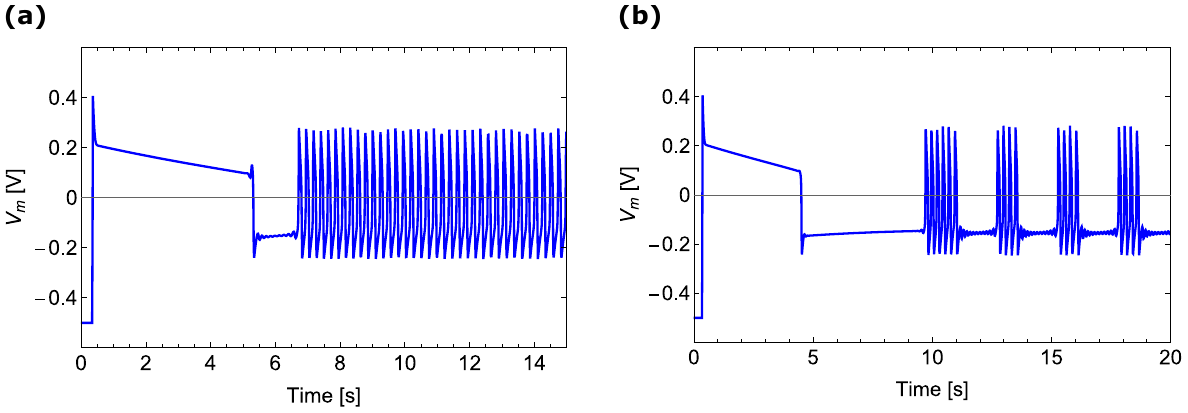}
		\caption{Two modes of voltage spiking extracted by modeling one and the same iontronic circuit with UP channels driven by different time-dependent currents, with \textbf{(a)} tonic spiking, i.e.\ regular spiking, for a stimulus of $1.495$ pA, and \textbf{(b)} tonic bursting,  i.e. \ a short burst of spiking alternating with periods of quiescence, for a stimulus of $1.485$ pA. System parameters are given in the text.}
		\label{fig:SI_Fig}
\end{figure*}
Our predictions of the dynamic conductance of UP channels is based on an analytical model that describes their steady-state \cite{Boon2022Pressure-sensitiveGeometry} and dynamic \cite{Kamsma2023IontronicMemristors} conductance properties. UP channels exhibit a voltage-dependent concentration polarisation that can reasonably accurately be described analytically by a slab-averaged salt concentration profile
\begin{equation}\label{eq:rhosProf}
\begin{aligned}
    \overline{\rho}_{i,\mathrm{\mathrm{s}}}&(x,V_i)\\
    &=2\rho_{\mathrm{b}}+2\rho_{\mathrm{b}}\Delta g\left[\frac{x}{L_i}\frac{R_{\mathrm{t}}}{R_i(x)}-\frac{e^{\text{Pe}_i(V_i)\frac{x}{L_i}\frac{ R_{\mathrm{t}}^2}{R_{\mathrm{b}}R_i(x)}}-1}{e^{\text{Pe}_i(V_i)\frac{R_{\mathrm{t}}}{R_{\mathrm{b}}}}-1}\right],
\end{aligned}
\end{equation}
where $\Delta g\equiv-2w(\Delta R/R_{\mathrm{b}})\text{Du}$ with $w=eD\eta/(k_{\rm{B}}T\epsilon \psi_0)$ the ratio of ionic to electro-osmotic mobility \cite{Aarts2023IontronicDiscussion} and the tip Dukhin number $\text{Du}=\sigma/(2\rho_{\mathrm{b}}R_{\mathrm{t}})$. The temperature is set at $T=293.15$ K throughout. For the negative surface charge we consider here on the channel walls, Eq.~(\ref{eq:rhosProf}) describes an enhancement of the steady-state salt concentration in the channel at $V_i<0$, and a decrease at $V_i>0$.

Following Ref. \cite{Kamsma2023IontronicMemristors}, we approximate the static conductance $g_{i,\infty}(V_i)$ to follow from the salt concentration profile according to
\begin{equation}\label{eq:rhos}
    \begin{aligned}
    \frac{g_{i,\infty}(V)}{g_{i,0}}=&\int_0^{L_i} \overline{\rho}_{\mathrm{i,\mathrm{s}}}(x,V_i) \dd x/(2\rho_{\rm{b}}L_i).
    \end{aligned}
\end{equation}
This is a simplification compared to the more accurate dependence on $L_i/\int_{0}^{L_i}(\overline{\rho}_{i,\mathrm{s}}(x,V_i))^{-1} \dd x$ used in Eq.~(3) in the main text, although Eq.(\ref{eq:rhos}) was also found to still work reasonably well \cite{Boon2022Pressure-sensitiveGeometry,Kamsma2023IontronicMemristors}. Here $g_{i,0}=(\pi R_{\mathrm{t}} R_{\mathrm{b}}/L)(2\rho_{\rm{b}}e^2D/k_{\mathrm{B}}T)$ is the conductance of the UP channel in equilibrium. Eq.~(\ref{eq:rhos}) combined with Eq.~(\ref{eq:rhosProf}) predicts that the conical UP channel is a current rectifier since, for surface charge $\sigma<0$,  a negative voltage $V_i<0$ increases $g_{i,\infty}(V)$ while a positive applied voltage $V_i>0$ decreases $g_{i,\infty}(V)$, with respect to $g_{i,0}$. The difference is caused by salt accumulation and depletion in the channel for $V_i<0$ and $V_i>0$, respectively.

The typical time it takes for salt to accumulate or deplete, and hence the typical conductance memory retention time of the channel, is independent of the surface charge \cite{Kamsma2023IontronicMemristors} and identical to the time scale of the BP channels of the main text given by
\begin{align}
    \tau_i=\frac{L_i^2}{12D}.\label{eq:ts} 
\end{align}
The dynamic conductance $g_i(t)$ of channels of type $i$ is, similar to BP channels, well-described by the differential equation
\begin{equation}
	\dfrac{\dd g_i(t)}{\dd t}=\frac{g_{i,\infty}(V_i(t))-g_i(t)}{\tau_i},\label{eq:deqrho}
\end{equation}
where $V_i(t)$ is the time-dependent voltage drop between base and tip of the channel.

\section{Tonic bursting with unipolar channels}

To investigate the emergence of bursting behaviour we consider the same circuit as in the main text, but with UP channels, different circuit parameters and three super slow channels (each connected in series to individual batteries) connected in parallel, which is mathematically equivalent to tripling the conductance of a single super slow channel. The two fastest and shortest channels have an individual battery in series with potential $E_\pm=\pm0.975$ V, and are of length $L_\pm=\frac{8}{9}$ $\mu$m $\approx0.89$ $\mu$m, hence with timescale $\tau_\pm\approx0.038$ ms. The slow channel is connected in series to a battery with potential $E_s=-0.5$ V, with the channel length $L_s=27.75$ $\mu$m and thus a timescale $\tau_s\approx 37$ ms. Finally, the three super slow (and even longer) channels that are placed parallel of each other each have an individual battery in series with potential $E_{\mathrm{ss}}=-0.5$ V and a channel length of $L_{\mathrm{ss}}=277.5$ $\mu$m, resulting in a timescale of $\tau_{\mathrm{ss}}\approx3.7$ s. From now on, these three channels are denoted as a single super slow channel with base conductance $3g_{0,\mathrm{\mathrm{ss}}}$, which is mathematically equivalent since they are described by identical equations. The only additive value of three channels rather than one is their increased total conductance that is necessary for a chattering spiking pattern to emerge in this instance. The capacitor has capacitance $C=4$ fF and a stimulus current $I(t)$ can be imposed through the circuit. Via Kirchhoff's law we find the following equation to describe the time-evolution of the voltage $V_{\mathrm{m}}(t)$ over the capacitor,
\begin{align}
	C\dfrac{\dd V_{\mathrm{m}}(t)}{\dd t}=I(t)-\sum_{i}g_{i}(t)\left(V_{\mathrm{m}}(t)-E_{i}\right),\label{eq:Kirchh}
\end{align}
where $i\in\left\{+,-,\mathrm{\mathrm{s}},\mathrm{\mathrm{ss}}\right\}$ and with the voltage arguments $V_i(t)$ over the channels given by $V_-(t)=V_{\mathrm{m}}(t)-E_-$, $V_+(t)=-V_{\mathrm{m}}(t)+E_+$, $V_{\mathrm{\mathrm{s}}}(t)=-V_{\mathrm{m}}(t)+E_{\mathrm{\mathrm{s}}}$, and $V_{\mathrm{\mathrm{ss}}}(t)=-V_{\mathrm{m}}(t)+E_{\mathrm{\mathrm{ss}}}$.

By numerically solving for $V_{\mathrm{m}}(t)$ for sustained current stimuli that undergo a step from 0 pA to $1.495$ pA and from 0 pA to $1.485$ pA, we find the voltage responses shown in Fig.~\ref{fig:SI_Fig}(a) and Fig.~\ref{fig:SI_Fig}(b), respectively. In Fig.~\ref{fig:SI_Fig}(a) we see after a transient has passed upon applying the stimulus that the circuit settles into tonic spiking. In Fig.~\ref{fig:SI_Fig}(b) we show that tonic bursting appears after a similar transient as in Fig.~\ref{fig:SI_Fig}(a). Therefore we can also produce the additional spiking mode of tonic bursting with UP channels. However, the typical potentials during spiking are further removed from typical voltages in neurons \cite{Bean2007TheNeurons} compared to the results in the main text. Additionally, the higher battery potentials and lower salt concentration and capacitance in this instance are not as similar to their biological counterparts \cite{fundNeuroTrain,Gentet2000DirectNeurons,Major1994DetailedSlices} as the values used in the BP channel circuit discussed in the main text. Lastly, while the approach of connecting three super slow channels in parallel is in principle physical, it further complicates the circuit design. Nevertheless, the existence of tonic bursting in a similar circuit with different fluidic memristors does show that bursting spiking modes can be achieved using multiple types of iontronic devices, suggesting that further improvements can be achieved by using even more desirable devices than the BP channels used in the main text.


%